\documentclass[twoside,amsmath,amssymb,nofootinbib,superscriptaddress,notitlepage]{revtex4-1}
\usepackage{graphicx,subfigure}
\usepackage{dcolumn,marvosym}
\usepackage{float,hyperref}
\usepackage{empheq,cleveref}
\usepackage{verbatim,times}
\usepackage{color}
\usepackage{enumitem}
\usepackage{amsmath}
\usepackage{amsfonts}
\usepackage{graphicx,braket}
\usepackage{bookmark}
\begin{document}

\title{Holographic estimate of heavy quark diffusion in a magnetic field}
\author{David Dudal}
\email{david.dudal@kuleuven.be}
\affiliation{KU Leuven Campus Kortrijk -- Kulak, Department of Physics, Etienne Sabbelaan 53 bus 7657, 8500 Kortrijk, Belgium}
\affiliation{Ghent University, Department of Physics and Astronomy, Krijgslaan 281-S9, 9000 Gent, Belgium}
\author{Thomas G.~Mertens}
\email{thomas.mertens@ugent.be}
\affiliation{Ghent University, Department of Physics and Astronomy, Krijgslaan 281-S9, 9000 Gent, Belgium}
\affiliation{Joseph Henry Laboratories, Princeton University, Princeton, NJ 08544, USA}

\begin{abstract}
We study the influence of a background magnetic field on the $J/\psi$ vector meson in a DBI-extension of the soft wall model, building upon our earlier work \cite{Dudal:2014jfa}. In this specific holographic QCD model, we discuss the heavy quark number susceptibility and diffusion constants of charm quarks and their dependence on the magnetic field by either a hydrodynamic expansion or by numerically solving the differential equation. This allows us to determine the response of these transport coefficients to the magnetic field. The effects of the latter are considered both from a direct as indirect (medium) viewpoint. As expected, we find a magnetic field induced anisotropic diffusion, with a stronger diffusion in the longitudinal direction compared to the transversal one. We backup, at least qualitatively, our findings with a hanging string analysis of heavy quark diffusion in a magnetic field. From the quark number susceptibility we can extract an estimate for the effective deconfinement temperature in the heavy quark sector, reporting consistency with the phenomenon of inverse magnetic catalysis.
\end{abstract}

\maketitle
\bookmarksetup{startatroot}


\section{Introduction}
Brownian diffusion of a charged particle of mass $m$ and charge $q$ in a background magnetic field $\mathbf{B}$ can be modeled using the Langevin equation,
\begin{equation}
\label{lange}
\frac{d\mathbf{p}}{dt}=-\gamma \mathbf{p} + q\mathbf{v}\times\mathbf{B}+\mathbf{R}(t),
\end{equation}
where the first term describes friction, the second one the Lorentz force and the third one white Gaussian random noise,
\begin{equation}
\braket{R_i(t)}=0\,,\qquad \braket{R_i(t_1)R_j(t_2)}=\kappa \delta_{ij} \delta(t_1-t_2).
\end{equation}
This has been studied extensively in the non-equilibrium statistical mechanics literature, see e.g.~\cite{Taylor,Kursunoglu} for some classical works on the subject and \cite{Aquino,Lisy} for some more recent studies. Some interesting background information can also be found in \cite{Fischler:2012ff}. If we consider a constant magnetic field $\mathbf{B}=B\mathbf{e}_3$ along the 3rd spatial direction, the diffusion is only affected by the magnetic field along the transverse directions, yielding
\begin{equation}\label{lang1}
D_\|=\frac{T}{\gamma m}\,,\qquad D_\perp=\frac{D_\|}{1+\frac{q^2B^2}{m^2\gamma^2}}.
\end{equation}
In general, we expect the influence of the magnetic field to arise from two distinct sources. Firstly, there is a direct influence on the motion of the charged particle itself. This is modeled by inclusion in the Langevin equation of a deterministic Lorentz force due to the magnetic field, as shown in equation (\ref{lange}). Secondly, we expect the magnetic field to influence the thermal background, which in turn can influence the diffusion of the charged particle. This effect is indirect, and as a first order approximation, we may imagine we can neglect this part. In principle, it should be modeled by recomputing the mean squared momentum transferred per unit time in a background magnetic field, by for instance perturbatively recomputing the scattering amplitudes \cite{Moore:2004tg}. This effect hence comes from microscopic interactions.

In this paper, we will focus our attention on the diffusion of a heavy quark, such as the charm $c$, in the quark-gluon plasma, in particular we are interested in how a magnetic field can influence this diffusion. This is of interest to understand how charmonia as $J/\Psi$, a heavy $c$ and $\bar c$ bound state, react to the presence of a strong magnetic field. QCD supplemented with a classical magnetic background attracted a great deal of interest over the past decade, given the expectation that such a field is generated during a non-central heavy ion collision and persists long enough to also influence the generated quark-gluon plasma \cite{Kharzeev:2007jp,Skokov:2009qp,Bzdak:2011yy,Deng:2012pc,Tuchin:2013apa,Tuchin:2013ie,McLerran:2013hla}. The relevance of charmonia dissocation (melting) in relation to the quark-gluon plasma creation is well-appreciated since the standard work \cite{Matsui:1986dk}. A renowned lattice related study is \cite{Asakawa:2003re}. Due to the strongly coupled nature of the problem under study, we will rely on the gauge-gravity paradigm, and consider a holographic version of QCD that includes charmonia dynamics. Other holographic works on charmonia can be found in \cite{Peeters:2006iu,Kim:2007rt,Hou:2007uk,Fujita:2009wc,Fujita:2009ca,Grigoryan:2010pj,Hohler:2013vca,Ali-Akbari:2014vpa,Ali-Akbari:2015bha,Braga:2016wkm,
Braga:2016oem,Braga:2017bml}, inclusion of magnetic field effects on charmonia dynamics was first considered in our previous paper \cite{Dudal:2014jfa}. Other takes on this particular problem can be found in e.g.~\cite{Machado:2013rta,Alford:2013jva,Cho:2014loa,Bonati:2015dka,Sadofyev:2015hxa,Suzuki:2016kcs,Yoshida:2016xgm,Hattori:2016emy,Suzuki:2016fof}, while more generally the response of the QCD phase transitions to a magnetic field have been intensively discussed in a pleiad of papers, a selection thereof is
\cite{Gusynin:1994re,Miransky:2002rp,Mizher:2010zb,Fraga:2008um,Gatto:2010pt,Gatto:2010qs,Osipov:2007je,Kashiwa:2011js,Fraga:2013ova,Filev:2007gb,Albash:2007bk,Evans:2010xs,Alam:2012fw,Preis:2010cq,
Filev:2010pm,Callebaut:2011zz,Callebaut:2013ria,Ballon-Bayona:2013cta,Bolognesi:2011un,Bali:2011qj,Bali:2012zg,Bali:2014kia,Bonati:2017uvz,Bonati:2013lca,D'Elia:2011zu,D'Elia:2010nq,Ilgenfritz:2013ara,
Ilgenfritz:2012fw,Frasca:2011zn,Fukushima:2012xw,Ferreira:2013oda,McInnes:2015kec,Dudal:2015wfn,Dudal:2016axr,Dudal:2016joz,Critelli:2016cvq,Rougemont:2015oea,Mueller:2015fka,Ayala:2014iba,Gursoy:2016ofp,Rodrigues:2017iqi,Fraga:2012ev,Fraga:2012fs,Ferreira:2014kpa,Fukushima:2012kc,Ayala:2014gwa,Farias:2014eca,Costa:2015bza,Mamo:2015dea,Fang:2016cnt,Li:2016gtz,Li:2016gfn,Li:2016smq,Kharzeev:2012ph,Miransky:2015ava}.

The diffusion in a magnetic field leads to conceptual complications, as the introduction of $\mathbf{B}$ leads to an anisotropic behavior, already visible from the classical result \eqref{lang1}. In \cite{Fukushima:2015wck,Hattori:2016idp}, the transverse and longitudinal momentum diffusion coefficients $\kappa_\|$ and $\kappa_\perp$ were computed in a modified hard thermal loop expansion (valid for $\alpha_s eB \ll T^2 \ll eB$), with as estimate for their relative behavior,
\begin{eqnarray}\label{f1}
\frac{\kappa_\|}{\kappa_\perp}&\sim& \frac{T^2}{eB}\ll 1.
\end{eqnarray}
This sizeable anisotropy could be an important ingredient in quantifying the elliptic flow (the out-of-reaction-plane anisotropy in particle production) \cite{Adare:2006nq,Adare:2010de,Abelev:2013lca,Scardina:2015fxa}, as explained in \cite{Muller:2013ila,Fukushima:2015wck,Hattori:2016idp}. More light on this could be shed from (hydrodynamic) simulations of the evolution of the quark-gluon plasma, \cite{Shuryak:2008eq}, using as input reliable estimates for the transport coefficients as $\kappa_\|$ and $\kappa_\perp$. Hence our interest to investigate these transport coefficients of the quark-gluon plasma. Other transport coefficients were also considered in e.g.~\cite{Hattori:2017qih,Li:2017tgi}.

Using the holographic charmonium model in our previous work \cite{Dudal:2014jfa}, we already obtained
\begin{eqnarray}\label{f2}
\frac{D_\|}{D_\perp}&\sim& 1+ C \frac{(eB)^2}{T^4},
\end{eqnarray}
with $C\approx 4.95$  for the diffusion coefficients in position space. In the absence of anisotropy, one has $D=\frac{2T^2}{\kappa}$ by means of the fluctuation-dissipation theorem, which is essentially a consequence of equipartition, stating that $\braket{p_i^2}=mT$, for each $i$. Assuming a Brownian motion in magnetic field described by \eqref{lange}, we a priori have only a single $\kappa$, which can be related to the two position space diffusion constants via
\begin{eqnarray}\label{f3}
D_\|=\frac{\kappa}{2m^2}\,,\qquad D_\perp=\frac{\kappa}{2m^2\left(\gamma^2+\frac{q^2B^2}{m^2}\right)}
\end{eqnarray}
which is just a rewriting of \eqref{lang1} given that $\gamma=\frac{\kappa}{2mT}$. The same equipartition still applies, see e.g.~\cite{Aquino,Fischler:2012ff}. Though, we shall refrain from directly studying the momentum diffusion constants in this paper.\footnote{Notice that \cite{Fukushima:2015wck} use a different anisotropic Langevin equation, not including the Lorentz force.} A detailed analysis of the different transport coefficients, including friction, not using any a priori Langevin connection with the spatial diffusion constants,  would entail the (not necessarily) spectral study of the full set of transport coefficients \cite{Gubser:2006bz,CasalderreySolana:2006rq,CaronHuot:2009uh,Finazzo:2016mhm,Zhang:2018mqt}. A general analysis of diffusion in anisotropic backgrounds was provided in \cite{Giataganas:2013hwa,Giataganas:2013zaa,Giataganas:2018ekx}.

Our estimate \eqref{f2} can be seen as complementary to the result \eqref{f1}, since \eqref{f2} is by construction obtained at strong coupling, the natural playground of AdS/QCD. We did not determine yet the separate quantities $D_\|$ and $D_\perp$ in \cite{Dudal:2014jfa}, a situation that will be remedied in the current work.

Returning to the Brownian motion, it must be mentioned here to what extent it is difficult to model the indirect magnetic effect at the level of the quark-gluon plasma. Modeling a heavy quark in QCD in a holographic fashion is easiest done in bottom-up models. It is known how to include heavy quarks using a variety of D-brane constructions, but these are usually in models not directly equivalent to QCD, or are rather complicated to work with as the metric is determined purely numerically, for a relevant example see \cite{Gursoy:2016ofp}. A simplified way, still capable of capturing essentials of QCD, is to take a phenomenological model, such as the celebrated soft wall model \cite{Karch:2006pv,Herzog:2006ra}, and add heavy quarks to it.

Within the same phenomenological route, the heavy quark current is dual to a bulk gauge field. The influence of the magnetic field can be dealt with in several ways (which are not equivalent). The simplest one, is to directly include the electromagnetic coupling to the heavy quark itself. We modeled this in \cite{Dudal:2014jfa} by using a DBI action instead of the usual Maxwell action, without it originating necessarily from an underlying brane system. Let us briefly reiterate the underlying motivation for the introduction of an effective DBI description. As we wish to couple a magnetic field to the electromagnetically neutral $J/\Psi$, this can only happen when the field couples to the charged constituents of the charmonium, i.e.~we must probe the internal structure of the meson. As a 5D Maxwell action treats the $J/\Psi$ as a point particle, we will rather rely on a 5D DBI generalization, keeping in mind this was also the original motivation why the non-linear DBI electrodynamics was introduced: to smear out point-like charges, this to overcome the infinite self-energy. Secondly, the magnetic field is expected to modify the thermal medium as well, which is neglected here at first, as in the early diffusion literature. This corresponds to only considering gluons in the thermal gas, the so-called quenched approximation.  Later on, we will also include medium effects by using the DBI extended soft wall model, but in a magnetized AdS background rather than in the usual AdS background, following the works \cite{D'Hoker:2009mm,D'Hoker:2009ix,Dudal:2015wfn}.

To sum up, not many models would be capable of studying heavy quarks in QCD in magnetic fields in the strongly coupled thermal medium, and we are naturally led to see what our model does, and how it gives different results compared to the superconformal $\mathcal{N}=4$ Supersymmetric Yang-Mills (SYM) case.
We pay special attention to comparing our results with multiple alternative approaches: the free-field approximation, Langevin results, $\mathcal{N}=4$ computations and a hanging string method.

The paper is organized as follows. Section~\ref{sect2} summarizes the heavy quark extension of the soft wall model by a DBI term, to incorporate a magnetic field coupling to the charmonium state, as first introduced in \cite{Dudal:2014jfa}. The quark number susceptibility, both from an analytical hydrodynamic or numerical differential equation viewpoint, is discussed in Section~\ref{sect4}, where we also provide an estimate for the deconfinement temperature. The diffusion constants themselves are derived and discussed in Section~\ref{sect5} from a spectral function viewpoint. Section~\ref{sect6} reanalyzes the diffusive behavior from a more stringy viewpoint, finding qualitative agreement with the spectral function analysis. In Section~\ref{sect6b} we generalize our model to include the indirect effects of the magnetic field on the medium, by utilizing a magnetic field dependent metric. We compare direct and indirect effects. Finally, Section~\ref{sect7} contains our conclusions and outlook for future work.

\section{The holographic charmonium model in a magnetic field}\label{sect2}
We review here our proposed bottom-up holographic model to describe charmonium physics \cite{Dudal:2014jfa}. \\
Let us start by giving the relevant backgrounds used in the original soft wall model. Our model can be viewed as a modification of the original soft wall model, which is capable of probing the internal charged structure of mesons. We consider two spacetimes relevant for our purposes. The first is AdS$_5$ (in Poincar\'e coordinates):
\begin{align}
\label{ads}
ds^2 &= \frac{L^2}{z^2}\left(-dt^2 + d\mathbf{x}^2 + dz^2\right)\,,\qquad e^{-\Phi} = e^{-c_\rho z^2},
\end{align}
giving the low temperature confined phase, as discussed in \cite{Karch:2006pv} in terms of a QCD-like confined spectrum at $T=0$, obtained from considering a 5D flavour action in this particular background. At $T>0$, it is tacitly assumed that the (Euclidean) temporal direction is compactified as usual with radius corresponding to the inverse temperature. The $z$-coordinate ranges from 0 (the boundary) to $\infty$ (the AdS deep interior). The second is the (planar) AdS black hole
\begin{align}
ds^2 &= \frac{L^2}{z^2}\left(-f(z)dt^2 + d\mathbf{x}^2 + \frac{dz^2}{f(z)}\right) \,,\qquad
e^{-\Phi} = e^{-c_\rho z^2},
\end{align}
representing the high temperature deconfined phase, where $f(z) = 1 - \frac{z^4}{z_h^4}$. In this case, the radial $z$ coordinate ranges from 0 to $z_h$, the horizon of the black hole. The temperature of the boundary theory equals the Hawking temperature of the black hole: $T = \frac{1}{\pi z_h}$. The scale is set by the parameter $c_\rho=0.151~\text{GeV}^2$ so that the ground state mass of the $\rho$-meson, $m_\rho=0.770~\text{GeV}$, is reproduced \cite{Herzog:2006ra}. There is a deconfinement transition at $T_c=0.191~\text{GeV}$, corresponding to a Hawking--Page transition between both above spacetimes \cite{Herzog:2006ra}, which can be related to a different behaviour of the Polyakov loop in the dual field theory, as expected from the viewpoint of a confined vs.~deconfined phase.

The soft wall in the AdS bulk modifies the action in the deep interior, and allows a better matching to actual QCD. The price one pays here is that Einstein's equations are not satisfied. It is possible to remedy this, by necessarily complicating the model, but we follow the original spirit of \cite{Karch:2006pv} and view it as a phenomenological toy model to develop a first intuition in the relevant physics without having to delve in numerous numerical details. Within the above space-time, one can now consider gauge fluctuations, dual to vector currents on the boundary theory as:\footnote{The normalization of this action has been determined in \cite{Colangelo:2008us} by matching with QCD flavor-flavor correlators. The gauge field considered here is in the vector subalgebra: $V = \frac{A_L + A_R}{2}$ of the full gauge sector. This normalization will be important further on when we compare to free-field results.}
\begin{equation}
\label{softwall}
S = - \frac{N_c}{24\pi^2 L}\int d^{5}x \sqrt{-g} e^{-c z^2} \text{tr}F^{\mu\nu}F_{\mu\nu},
\end{equation}
for a $SU(N_f)$ flavor group. Following \cite{Fujita:2009wc,Fujita:2009ca}, we use this action to describe the heavy charm vector current (the $J/\psi$ mesons), where $c = 2.43$ GeV$^2$, obtained by matching the $J/\psi$ spectrum with experiment. Hence we reduce to a $U(1)$ flavor group with $F = dV$ for a gauge field $V$. The relative size of $c$ compared to $c_\rho$, viz.~the improved exponential dampening in the action, renders the charm contribution essentially irrelevant for the numerical determination of the aforementioned deconfinement temperature $T_c$. Where appropriate, we will work with a dimensionless temperature
\begin{equation}
\label{temp}
t=\frac{T}{\sqrt c}.
\end{equation}
Doing so, we also have the critical $t_c=0.12$.

However, as pointed out in \cite{Dudal:2014jfa}, the action \eqref{softwall} misses essential physics when turning on external magnetic fields, modeled in through the flavor-diagonal part of $F_{\mu\nu}$. The reason is the linearity of the Maxwell fields: the fluctuations do not feel the magnetic field itself. At lowest order, this makes sense as the vector field corresponds to the $c\bar{c}$ state which has zero charge. But one does expect a nontrivial influence from its internal quark charges (dipole moment). This led us to propose a simple DBI extension to describe $J/\psi$ mesons interacting with external electromagnetic fields as:\footnote{We will discuss our strategy to fix the string length $\ell_s=\sqrt{\alpha'}$ later on.}
\begin{equation}
\label{DBI}
S = -\frac{N_c}{24\pi^4 \alpha'^2 L}\int d^{5}x e^{-\Phi}\sqrt{-\det\left(g_{\mu\nu}+2\pi\alpha' i F_{\mu\nu}\right)}.
\end{equation}
We showed in \cite{Dudal:2014jfa} that the boundary Green functions one computes with this holographic action give reasonable spectral functions and locations of the bound state poles. We also initiated a study of the heavy quark diffusion coefficient, by taking the $\omega \to 0$ limit of a certain Green function
\begin{equation}
\label{Dform}
D = - \frac{1}{3\chi}\text{lim}_{\omega\to 0}\sum_{i=1}^{3}\frac{\Im G^{R}_{ii}}{\omega},
\end{equation}
with $\chi$ the quark number susceptibility. This expression for $D$ is a typical example of a Kubo relation, and it can be obtained from a linear response analysis: one departs from the underlying (conserved) heavy quark current and assumes a small perturbation away from equilibrium and carries out a gradient expansion of the current inside the current-current correlation function. A detailed derivation can be found in \cite{Laine:2016hma, pasztor}. The essentials have been summarized in Appendix \ref{formulas}, including the required modification in the anisotropic case, which is due to the anisotropy introduced by a background magnetic field. We need to split the diffusion constants into $D_{\perp}=D_1=D_2$ and $D_{\parallel}=D_3$. The required modification of (\ref{Dform}) in the anisotropic case can be found in \eqref{A5}.

The quantity $\chi$ itself was not yet determined in our earlier work, and is expected to also depend on the magnetic field. By taking the ratio of $D_{\perp}$ and $D_{\parallel}$, we already noticed in \cite{Dudal:2014jfa} that charm quarks diffuse much more easily in directions parallel to the magnetic field than in perpendicular directions. This confirms from a strong coupling analysis the results presented in \cite{Fukushima:2015wck,Hattori:2016idp}.

Our goal here is to progress further this story, by first explicitly computing $\chi$ itself holographically, and then analyzing $D_{\perp}$ and $D_{\parallel}$ as a function of the background magnetic field, comparing with a Langevin analysis and earlier SYM results whenever possible. From the novel result for $\chi$ we can also extract an estimate for the deconfinement temperature in the heavy quark sector as a function of the magnetic field.

The coordinates are denoted as $t, x_1, x_2, x_3$ for the dual coordinates and $z$ for the holographic coordinate. The background gauge field takes into account the magnetic field along the 3-axis as
\begin{equation}
\bar{F}_{12} = - \bar{F}_{21} = \partial_1 A_2 = -iqB\frac{2}{3}.
\end{equation}
The equations of motion following from the action (\ref{DBI}), expanded up to second order in the fluctuations, are
\begin{equation}
\label{master}
\partial_{\mu}\left(e^{-\Phi}\sqrt{-\mathcal{G}}F^{\mu\nu}\right)=0,
\end{equation}
where we defined
\begin{eqnarray}
\mathcal{G}_{\mu\nu}=\left[\begin{array}{ccccc}
g_{00} & 0 & 0 & 0 & 0\\
0 & g_{11} & 2\pi\alpha' i \bar{F}_{12} & 0 & 0  \\
0 & -2\pi\alpha' i \bar{F}_{12} & g_{22} & 0 & 0\\
0 & 0 & 0 & g_{33} & 0 \\
0 & 0 & 0 & 0 & g_{zz} \end{array}\right].
\end{eqnarray}
We will denote by $G$ only the symmetric part of the metric tensor $\mathcal{G}$, and indices of (\ref{master}) are raised and lowered with this tensor.
For later reference, its determinant equals
\begin{equation}
\mathcal{G} = g_{00}g_{33}g_{zz}\left(g_{11}g_{22} - (2\pi\alpha')^2\bar{F}_{12}^{2}\right),
\end{equation}
and its inverse is
\begin{eqnarray}
\mathcal{G}^{\mu\nu}=\left[\begin{array}{ccccc}
\frac{1}{g_{00}} & 0 & 0 & 0 & 0\\
0 & \frac{g_{22}}{X} & - \frac{2\pi\alpha' i \bar{F}_{12}}{X} & 0 & 0  \\
0 & \frac{2\pi\alpha' i \bar{F}_{12}}{X} & \frac{g_{11}}{X} & 0 & 0\\
0 & 0 & 0 & \frac{1}{g_{33}} & 0 \\
0 & 0 & 0 & 0 & \frac{1}{g_{zz}} \end{array}\right],
\end{eqnarray}
where $X = g_{11}g_{22} - (2\pi\alpha')^2\bar{F}_{12}^{2}$.

We choose the radial gauge $V_{z} = 0$ and focus on Fourier modes~$\sim e^{i\mathbf{k}\cdot \mathbf{x} - i \omega t}$ in the boundary theory.\footnote{General spectral functions in this model at finite momentum, along with a possible pitfall in the gauge choice for the black hole background were discussed in \cite{Dudal:2015kza}. }

\section{Quark number susceptibility from the spectral function approach}\label{sect4}
To find the separate diffusion coefficients, we will first need to determine the quark number susceptibility $\chi$ in the thermal plasma.
\subsection{Review: the quark number susceptibility in terms of the spectral function}\label{sect1}
First we review how $\chi$ is rewritten in terms of data obtained from the Green functions, following \cite{Kunihiro:1991qu}.
The quark number susceptibility $\chi$ is defined as the response of the quark number density $n$ to a change in chemical potential $\mu$ \footnote{To avoid confusion, we will work at vanishing density, so at the end, we ought to set $\mu=0$.}:
\begin{equation}
\chi = \left.\frac{\partial n}{\partial \mu}\right|_{\mu=0}.
\end{equation}
This quantity can be conveniently rewritten in terms of data obtainable from Green functions \cite{Kunihiro:1991qu} as follows.
In the grand canonical ensemble that we are considering, the quark number density is given by
\begin{equation}
nV = \text{Tr}\left(N\exp\left(-\beta(H-\mu N)\right)\right)/\mathcal{Z},
\end{equation}
where $N = \int dV J_0(t,\mathbf{x})$ with $J_0 = \bar{\psi}\gamma_{0}\psi$ and $\psi$ the quark field of interest.
One can now rewrite this in a first step as
\begin{equation}
\chi = \frac{\beta}{V} \left\langle N^2\right\rangle_{\beta}.
\end{equation}
Using translational invariance to cancel one spatial integral, one can rewrite this as
\begin{equation}
\chi = \beta \int dV\left\langle  J_0(0,\mathbf{x}) J_0(0,\mathbf{0})\right\rangle_{\beta},
\end{equation}
or upon Fourier transforming
\begin{equation}
\chi = \beta \lim_{\mathbf{k} \to \mathbf{0}} \int_{-\infty}^{+\infty}\frac{d\omega}{2\pi}S_{00}(\omega,\mathbf{k}),
\end{equation}
where
\begin{equation}
S_{00}(\omega, \mathbf{k}) = \int dt \int dV e^{i\mathbf{k}\cdot \mathbf{x}}e^{-i\omega t}\left\langle J_{0}(t,\mathbf{x})J_{0}(0,\mathbf{0})\right\rangle_{\beta}
\end{equation}
and $J_0 = \bar{\psi}\gamma_{0}\psi$.

Textbook arguments (see e.g.~\cite[Ch.~6]{Kapusta:2006pm}) now relate this two-point function to the imaginary part of the retarded Green function as
\begin{equation}
\Im G^{R}_{00}(\omega, \mathbf{k}) = - \frac{1-e^{-\beta \omega}}{2} S_{00}(\omega, \mathbf{k}),
\end{equation}
where
\begin{equation}
G^{R}_{00}(\omega, \mathbf{k}) = -i \int dt \int dV e^{i\mathbf{k}\cdot \mathbf{x}}e^{-i\omega t}\theta(t)\left\langle \left[J_{0}(t,\mathbf{x}),J_{0}(0,\mathbf{0})\right]\right\rangle_{\beta}.
\end{equation}
The retarded Green function (being causal) satisfies the Kramers-Kronig identity:
\begin{equation}
\Re G^{R}_{00}(\omega, \mathbf{k}) = \mathcal{P}\int\frac{d\omega'}{\pi}\frac{\Im G^{R}_{00}(\omega', \mathbf{k})}{\omega'-\omega}.
\end{equation}
Using now finally that
\begin{equation}
\lim_{\mathbf{k} \to \mathbf{0}} \Im G^{R}_{00}(\omega, \mathbf{k}) = - \frac{1-e^{-\beta \omega}}{2} 2\pi \delta(\omega)\frac{\chi}{\beta},
\end{equation}
\begin{equation}
\label{qnsus}
\chi = -\lim_{\mathbf{k} \to \mathbf{0}}\Re G^{R}_{00}(0, \mathbf{k}),
\end{equation}
relating the quark number susceptibility $\chi$ to a property of the retarded Green function of the quark currents (charm quarks in the case of interest).

Let us now discuss a few relevant examples.

At very high temperature, this quantity can be explicitly computed using asymptotic freedom, effectively leading to the expressions for a free quark-antiquark gas:
\begin{equation}
n = 2N_c\int \frac{d^3\mathbf{p}}{(2\pi)^3}\left(\frac{1}{e^{\beta(E-\mu)}+1} - \frac{1}{e^{\beta(E+\mu)}+1}\right), \quad E^2 = m^2 + \mathbf{p}^2.
\end{equation}
From this, one computes explicitly for $\mu=0$ and at high temperatures where the quark mass is negligible:
\begin{equation}
\label{chihight}
\chi = \left.\frac{\partial n}{\partial \mu}\right|_{\mu=0} = \frac{N_c}{3}T^2.
\end{equation}

For non-zero magnetic field, one needs the Landau levels as single-particle energies:
\begin{equation}
\label{landau}
E^2 = m^2 + p_z^2 + (2n-2s_z+1)|q|B,
\end{equation}
where $s_z = \pm \frac{1}{2}$ is the spin projection along the $B$-field.
Summing over spin projections, and including both quark and antiquark matter, this leads to
\begin{equation}
\label{chifreee}
\chi(B) = \left.\frac{\partial n}{\partial \mu}\right|_{\mu=0} = \frac{2N_c}{T} \int_{-\infty}^{+\infty}\frac{dp_z}{2\pi} \frac{|q|B}{2\pi} \sum_{n=0}^{+\infty}\left(2-\delta_{n,0}\right)\frac{e^{\frac{\sqrt{p_z^2+m^2+2n|q|B}}{T}}}{\left(e^{\frac{\sqrt{p_z^2+m^2+2n|q|B}}{T}}+1\right)^2}.
\end{equation}
This expression can also be derived by taking the second $\mu$-derivative, at $\mu=0$, of the free energy, as can be easily checked from the results in \cite{Preis:2012fh}. \\
Some simple checks on this formula \eqref{chifreee}, are that when $T\to +\infty$, one can approximate the summation over $n$ as an integral and this reduces to the free particle where $n \to \frac{p_x^2+p_y^2}{2|q|B}$.\footnote{Taking $|q|B \to 0$ in (\ref{landau}) allows for $n\to+\infty$ states surviving this limit: these precisely give states with non-zero transverse momenta $p_x$ or $p_y$.} The high temperature limit remains the same as (\ref{chihight}).\footnote{This follows also by simple dimensional considerations: the high $T$ limit is equivalent to the simultaneous $B,m\to 0$ limit.}\\
When $T\to 0$, the expression is dominated by the lowest Landau level $n=0$. After using the asymptotic series (for small $T$):
\begin{equation}
\int_{-\infty}^{+\infty}\frac{dp_z}{2\pi}\frac{e^{\frac{\sqrt{p_z^2+m^2}}{T}}}{\left(e^{\frac{\sqrt{p_z^2+m^2}}{T}}+1\right)^2} \, \approx \, 2\sqrt{\frac{\pi m}{2T}}e^{-\frac{m}{T}},
\end{equation}
one finds
\begin{equation}
\frac{\chi}{T^2} \sim \sqrt{\frac{m}{T}}\frac{|q|Be^{-\frac{m}{T}}}{T^2} \to 0.
\end{equation}
Note that for a massless particle, the limit works differently, and
\begin{equation}
\frac{\chi}{T^2} \sim \frac{|q|B}{T^2} \to +\infty,
\end{equation}
diverging for non-zero magnetic field in the low-temperature limit. \\
We will later on compare this free-field result to the holographic results.

\subsection{Computing the quark number susceptibility}
So, following (\ref{qnsus}) we are interested in computing the $00$ component of the retarded Green function for a generic (small) momentum dependence. \\
For the sake of simplicity, we take the momentum along the 3-axis, parallel to the external magnetic field. Since $\chi$ is a scalar quantity, this choice should be irrelevant in the end.\footnote{One readily checks that the hydrodynamic expansion method does not change the final expression for $V_0$ when we put $\mathbf{k}$ along the 1-axis for instance. The differential equation to be solved in that case is given by:
\begin{equation}
\partial^2_zV_0 + \partial_z\ln\left|e^{-\phi}\sqrt{-\mathcal{G}}G^{zz}G^{00}\right|\partial_z V_0 - \frac{G^{11}}{G^{zz}}k^2V_0 = 0,
\end{equation}
which does not alter the expression for $F_0$ (\ref{F0}) nor the fact that $V_0$ should vanish at the horizon $z=z_h$, as we explain further on.}
The relevant differential equation to be solved is equation (\ref{master}) in radial gauge $V_z=0$:
\begin{equation}
\label{ode1}
\partial^2_zV_0 + \partial_z\ln\left|e^{-\phi}\sqrt{-\mathcal{G}}G^{zz}G^{00}\right|\partial_z V_0 - \frac{G^{33}}{G^{zz}}k^2V_0 = 0.
\end{equation}
Our goal here will be to derive an analytic formula for $\chi$ (Section \ref{hydroexpp}) and compare it with the corresponding numerical solution (Section \ref{numsol}), reporting a perfect match.

\subsubsection{Solving the hydrodynamic expansion}
\label{hydroexpp}
Since we are interested in the small $k$ behavior, we can perform a hydrodynamic expansion:
\begin{equation}
V_0(z) =F_{0}(z) + k^2F_{k^2}(z) + \hdots,
\end{equation}
where terms of higher order in $\omega$ will be dropped from the start, since for $\chi$ we do not need them. \\

At lowest order, the differential equation (\ref{ode1}) reduces to:
\begin{equation}
\partial_z(X_{tz} \partial_z F_0) = 0,
\end{equation}
where
\begin{align}
X_{t3} &= e^{-\phi} \sqrt{-\mathcal{G}}G^{00}G^{33}, \\
X_{tz} &= e^{-\phi} \sqrt{-\mathcal{G}}G^{00}G^{zz}.
\end{align}
Integrating leads to
\begin{equation}
X_{tz}\partial_z F_0 = C_1.
\end{equation}
The third term in the differential equation (\ref{ode1}) has a simple pole at $z=z_h$. This requires $V_0(z_h) = 0$ at every order in $k^2$. Hence, we can integrate once more and obtain\footnote{Higher order terms in this series will not be needed, but can be determined. They all need to vanish at $z=z_h$ and are given explicitly by
\begin{equation}
F_{k^n}(z)=C_2 \int_{z_h}^{z}\frac{dz_{n+1}}{X_{tz}(z_{n+1})}\int_{z_h}^{z_{n+1}} X_{t3}(z_n)dz_n\int_{z_h}^{z_n} \frac{1}{X_{tz}(z_{n-1})}dz_{n-1}\int_{z_h}^{z_{n-2}} X_{t3}(z_{n-2})dz_{n-2}\hdots\int_{z_h}^{z_2} \frac{1}{X_{tz}(z_{1})}dz_{1}.
\end{equation}
These functions satisfy $F_{k^n}(z_h)=0$, $\partial_z F_{k^2}(z_h)\neq 0$, $\partial_z F_{k^n}(z_h) = 0$. The differential equation then implies that $V_0$ at higher orders in $k^2$ has a degenerate zero at $z=z_h$, whereas $V_0$ at lowest order has only a simple zero.}
\begin{equation}
\label{F0}
F_0(z) = C_1\int_{z_h}^{z}\frac{dz}{X_{tz}}.
\end{equation}

Finally, for $\chi$, one finds the closed expression\footnote{Note that this result is actually independent of the AdS length scale $L$, as it should be. We defined the quantity $\bar D=\frac{4\pi \alpha' qB}{3}$.}
\begin{equation}\label{chires}
\chi = \frac{N_c}{6\pi^2}\frac{1}{L\int_{0}^{z_h}\frac{dz}{e^{-cz^2}\sqrt{-\mathcal{G}}G^{00}G^{zz}}} = \frac{N_c}{6\pi^2} \frac{1}{L\int_{0}^{z_h}\frac{dz}{e^{-cz^2}\sqrt{\frac{L^{10}}{z^{10}}+\frac{L^6}{z^6}\bar D^2}\frac{z^4}{L^4}}},
\end{equation}
as a consequence of the real-time AdS/CFT correspondence set out in \cite{Son:2002sd,Policastro:2002se}, see also \cite{Teaney:2006nc}, which leads in our case to
\begin{equation}
\chi = \frac{N_c}{6\pi^2}\lim_{z\to0} \Re\left[\frac{1}{z}\frac{\partial_z V_0}{V_0}\right].
\end{equation}
For the necessary underlying details about the application of the real-time AdS/CFT PSS dictionary to our model, let us refer to \cite{Dudal:2014jfa}, as the steps leading to \eqref{chires} are rather analogous to those worked out there.

For $c=2.43$ GeV$^2$, the quark number susceptibility one finds has the shape shown in Figure~\ref{chiNint}. For the record, all shown quantities in this and the following figures are expressed in appropriate units of~$\text{GeV}$.

Note that we extended the figures also for temperatures below $t_c$ (or $T_c$) to gain intuition how the charmonia react to the full range of temperatures (modeled by the black hole geometry), although strictly speaking we should resort to the confined spacetime at lower temperatures.

\begin{figure}[h]
\begin{minipage}{0.32\textwidth}
\centering
\includegraphics[width=0.95\textwidth]{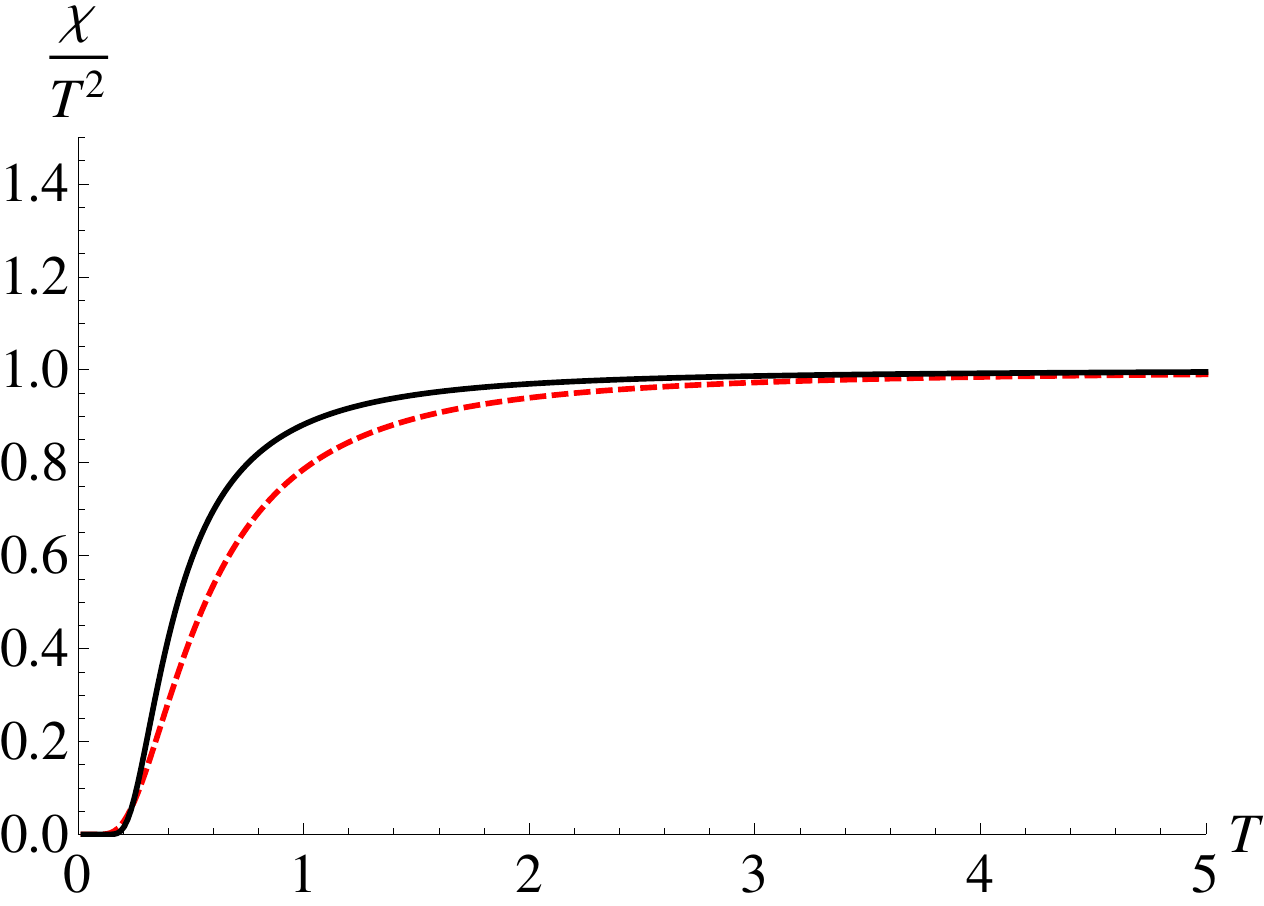}
\end{minipage}
\begin{minipage}{0.32\textwidth}
\centering
\includegraphics[width=0.95\textwidth]{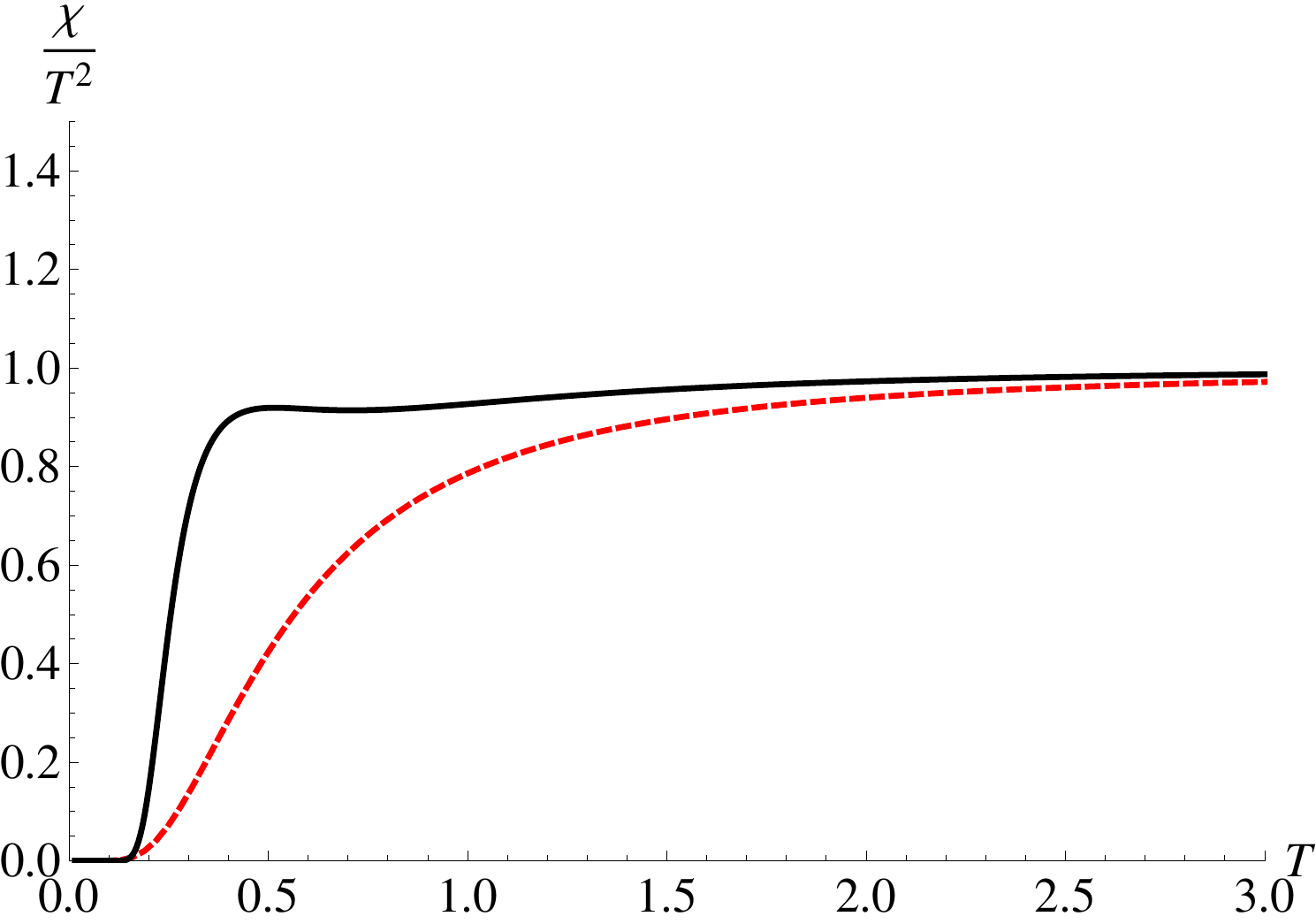}
\end{minipage}
\begin{minipage}{0.32\textwidth}
\centering
\includegraphics[width=0.95\textwidth]{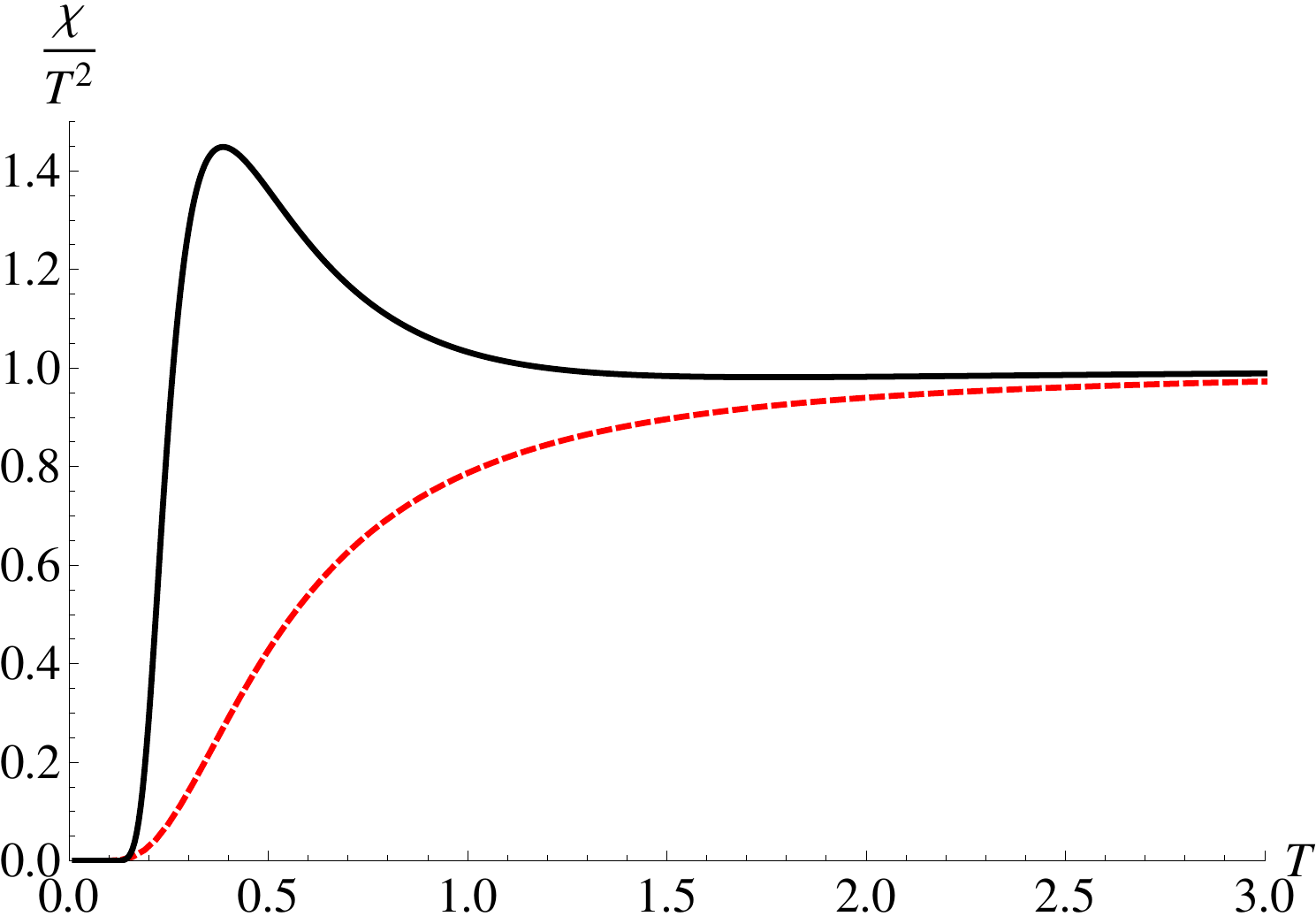}
\end{minipage}
\caption{Comparison of $\chi/T^2$ of the holographic result (full line) with the free fermion result of equation (\ref{chifreee}) (dashed line). Left Figure: $qB = 0$. Middle Figure: $qB = 0.25$ GeV$^2$. Right Figure: $qB = 0.5$ GeV$^2$.}
\label{compfree}
\end{figure}

Its behavior as a function of temperature $T$ for various values of $qB$ is shown in Figure~\ref{chivsT}. Note that for large values of $T$, the behavior is independent of $qB$, as expected indeed. The quark number susceptibility has the characteristic $\sim T^2$ behavior for large $T$ in the deconfinement regime.
\begin{figure}[h]
\begin{minipage}{0.48\textwidth}
\centering
\includegraphics[width=0.95\textwidth]{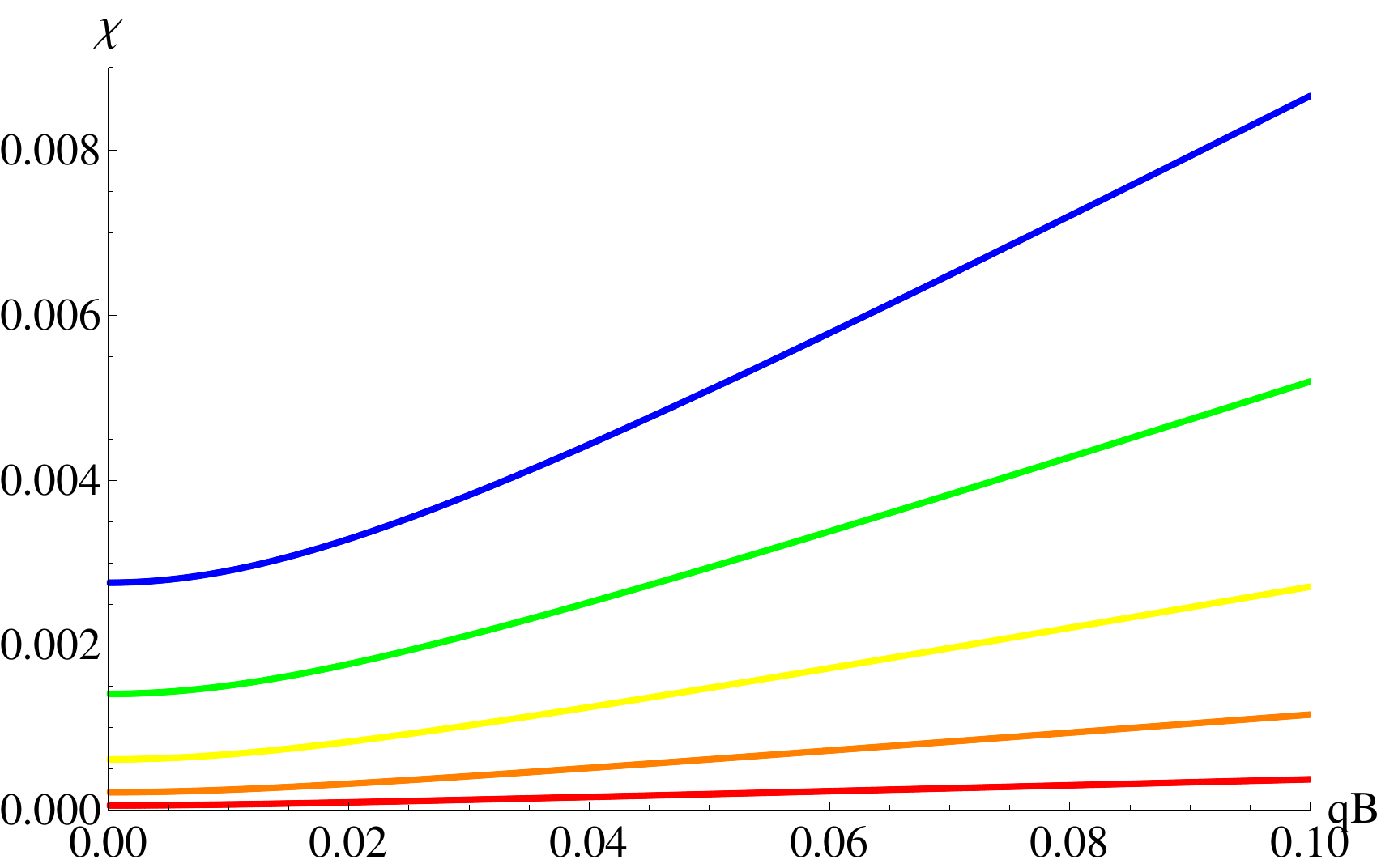}
\caption{Quark number susceptibility $\chi$ as a function of applied magnetic field. Red: $t=0.11$, yellow: $t=0.13$, green: $t=0.14$, blue: $t=0.15$.}
\label{chiNint}
\end{minipage}
\hfill
\begin{minipage}{0.48\textwidth}
\centering
\includegraphics[width=0.95\textwidth]{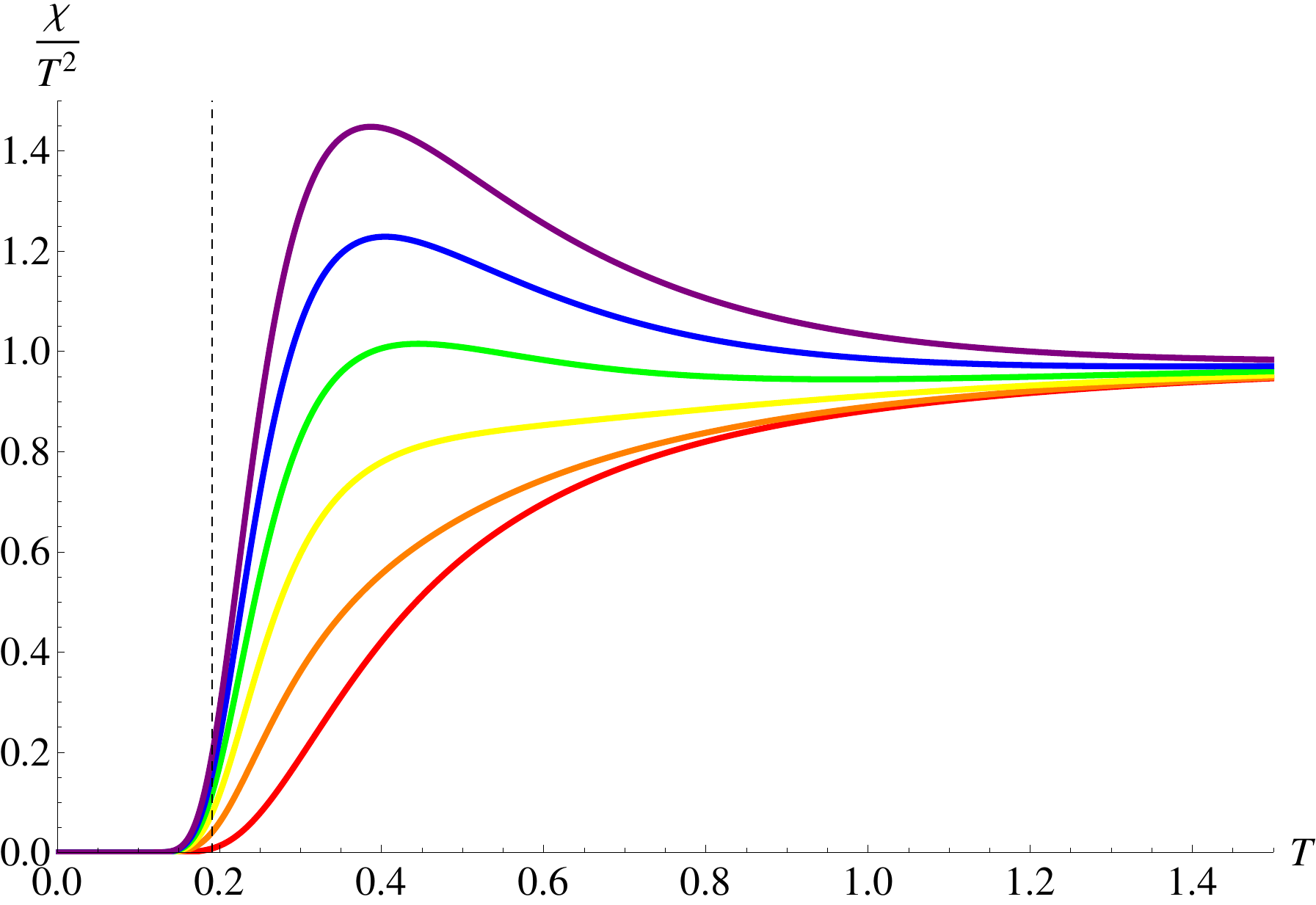}
\caption{Quark number susceptibility $\chi$ divided by $T^2$ as a function of temperature for various values of the magnetic field. Red: $qB=0$ GeV$^2$, orange: $qB= 0.1$ GeV$^2$, yellow: $qB= 0.2$ GeV$^2$, green: $qB= 0.3$ GeV$^2$, blue: $qB= 0.4$ GeV$^2$, purple: $qB= 0.5$ GeV$^2$. The dashed vertical line shows the deconfinement temperature $T_c=0.191$ GeV \cite{Herzog:2006ra} above which we should follow these curves. }
\label{chivsT}
\end{minipage}
\end{figure}
The soft wall model as it is, is incapable of producing a maximum of this curve around $T_c$. However, for larger values of the background magnetic field $qB$, a very pronounced maximum emerges automatically.
It is also interesting to compare the results given here to those of free fermions interacting with a background magnetic field \eqref{chifreee} (which should be a good approximation in the UV $T\to+\infty$ limit) (Figure~\ref{compfree}) \cite{Vuorinen:2002ue,Bazavov:2013uja}.

And indeed, the large $T$ region agrees with the free-field computations one finds for equation (\ref{chifreee}). We note that this agreement is a non-trivial consequence of correctly accounting for all of the prefactors in the soft-wall action (\ref{softwall}) we started with; it is a consistency check on the various normalizations.
From these Figures, it is also apparent that the characteristic peak we find holographically at higher values of $qB$ is absent in the free-field (UV) approximation.

For $qB=0$, one can perform the integral in (\ref{chires}) analytically: 
\begin{equation}
\chi = \frac{N_c}{6\pi^2} \frac{2c}{(e^{cz_h^2}-1)},
\end{equation}
agreeing with the result obtained earlier in \cite{Kim:2010zg}.

For large $qB$, one can approximate \eqref{chires} by\footnote{By neglecting the first term in the square root in the denominator compared to the second term. This is not fully trivial as the integral over $z$ ranges all the way to $0$ where the first term dominates. It turns out however that the largest contribution comes from the integration region close to $z_h$.}
\begin{equation}
\chi \approx -\frac{N_c}{6\pi^2} \frac{2}{\Re\,\text{Ei}(1,-cz_h^2)}\frac{4\pi\alpha' qB}{3L^2}.
\end{equation}
Linear behavior in $qB$ is apparent in Figure~\ref{chiNint} and the coefficient agrees with the one from the numerical integration displayed in Figure~\ref{chiNint} for large $qB$. Here, we have set $\ell_s = \sqrt{\alpha'} \approx 2.29L$, as it was fixed in our previous paper \cite{Dudal:2014jfa} by making a comparison with the lattice Polyakov loop. We will have to say more about the determination of $\alpha'$ later on.

\subsubsection{Numerical solution via the ordinary differential equation for $\omega=0$}
\label{numsol}
As an alternative to the hydrodynamic expansion, one can also numerically solve the differential equation. Here we present some of the details if one proceeds along this path, more can be found in \cite{Dudal:2014jfa}. In the current case, the differential equation to be solved is
\begin{equation}
\label{ode3}
\partial^2_zV_0 + \partial_z\ln\left|e^{-\phi}\sqrt{-\mathcal{G}}G^{zz}G^{00}\right|\partial_z V_0 - \frac{G^{33}}{G^{zz}}k^2V_0 = 0.
\end{equation}
A Frobenius analysis around the boundary $z=\epsilon$ shows that
\begin{align}
\Phi_1(\epsilon)&=1, \quad \Phi_1'(\epsilon) = k^2\epsilon\ln(\epsilon), \\
\Phi_2(\epsilon)&= \epsilon^2, \quad \Phi_2(\epsilon) = 2\epsilon.
\end{align}
Around the horizon $z=z_h$, one finds instead
\begin{align}
\Phi_1(\epsilon) &\sim \left(1-\frac{z}{z_h}\right), \\
\Phi_2(\epsilon) &\sim 1 + C\left(1-\frac{z}{z_h}\right)\ln\left(1-\frac{z}{z_h}\right).
\end{align}
Unlike the case when $\omega\neq0$, for which one imposes the real-time ingoing boundary condition at the horizon \cite{Son:2002sd,Policastro:2002se}, one cannot apply this condition in this case. Instead, as in the Euclidean case, we demand regularity at the horizon. While both solutions do not diverge themselves at the horizon, $\Phi_2$ has divergent derivatives and we discard it. From a different perspective, we argued in the previous Section that the correct solution should vanish at the horizon, requiring us to eliminate $\Phi_2$ indeed. \\

One readily finds agreement between the $\chi$ determined numerically by solving this ODE, and the value of $\chi$ found using the hydrodynamic expansion previously.

\subsection{Deconfinement transition from the quark number susceptibility}
\begin{figure}[h]
\centering
\includegraphics[width=0.4\textwidth]{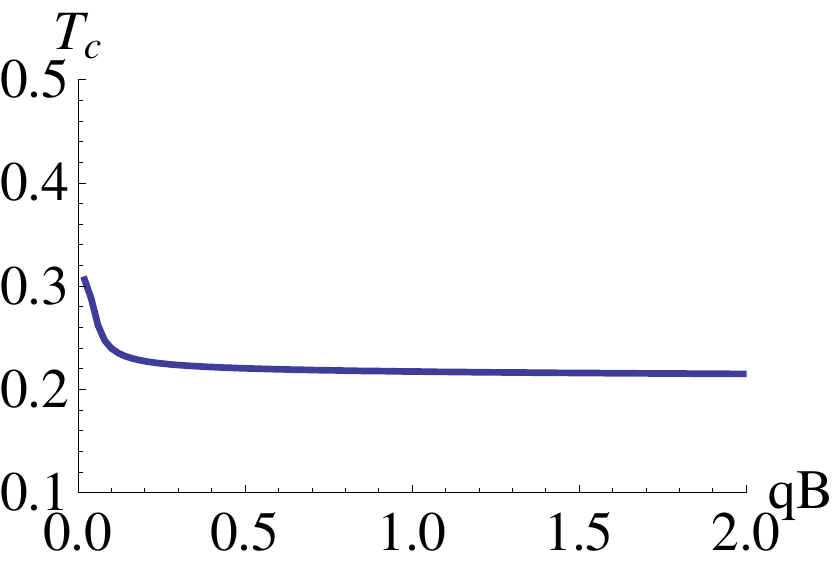}
\caption{Deconfinement temperature $T_c$ determined as the inflection point of the $\chi/T^2$ curve.}
\label{textr}
\end{figure}

Since one expects a rapid change in the fluctuations of the heavy quark number across deconfinement ($\sim$~the liberation of the charm quarks previously bound in heavy hadrons) \cite{Gottlieb:1987ac}, we can extract the critical temperature $T_{c}(B)$, following e.g.~\cite{Bazavov:2009zn,Bali:2011qj} by determining the inflection point of $\chi/T^2$ as a function of $T$. From Figure~\ref{chivsT} we already observe that $T_c(B)$ drops with magnetic field, in accordance with the lattice predictions of \cite{Bali:2011qj}. The numerics are displayed in Figure~\ref{textr}.

It is interesting to note that we find evidence for the inverse magnetic catalysis for $T_c$ using the original soft wall model, i.e.~without taking into account the backreaction of the magnetic field on the metric, which would correspond on the QCD side to the charged quarks coupling the magnetic field indirectly to the uncharged glue.

\section{Heavy quark diffusion from the spectral function approach}\label{sect5}
In our previous paper, \cite{Dudal:2014jfa}, we already computed the quantity $D\chi$ along and perpendicular to the magnetic field. We did this again both numerically (as a spin-off of analyzing the spectral functions) and analytically (using a hydrodynamic expansion). The results can be written down analytically as:
\begin{align}
\chi D_{\parallel} &= \frac{N_c}{6\pi}\frac{e^{-cz_h^2}}{z_h \pi}\sqrt{1 + \frac{16\pi^2(2.29)^4z_h^4}{9}(qB)^2}, \\
\label{Dperpe}
\chi D_{\perp} &= \frac{N_c}{6\pi}\frac{e^{-cz_h^2}}{z_h \pi}\frac{1}{\sqrt{1 + \frac{16\pi^2(2.29)^4z_h^4}{9}(qB)^2}}.
\end{align}
So, recuperating these results and dividing $\chi D$ by the newly obtained $\chi$, one finds the heavy quark diffusion coefficients, shown below in Figures~\ref{diff3Tb} and \ref{diff1Tb}.

We observe that for large $qB$ and for diffusion parallel to the applied magnetic field, $D$ becomes independent of $qB$, while perpendicular to the applied magnetic field, $D$ decays as $1/(qB)^2$ as $qB$ becomes large enough. This is in precise agreement with the analysis based on the Langevin equation which results in equation (\ref{lang1}).

For small values of $T$, the diffusion coefficient increases with $T$. This however changes at higher values of $T$ as shown explicitly below in Figures~\ref{diff3Tc} and \ref{diff1Tc}.
\begin{figure}[h]
\begin{minipage}{0.48\textwidth}
\centering
\includegraphics[width=0.95\linewidth]{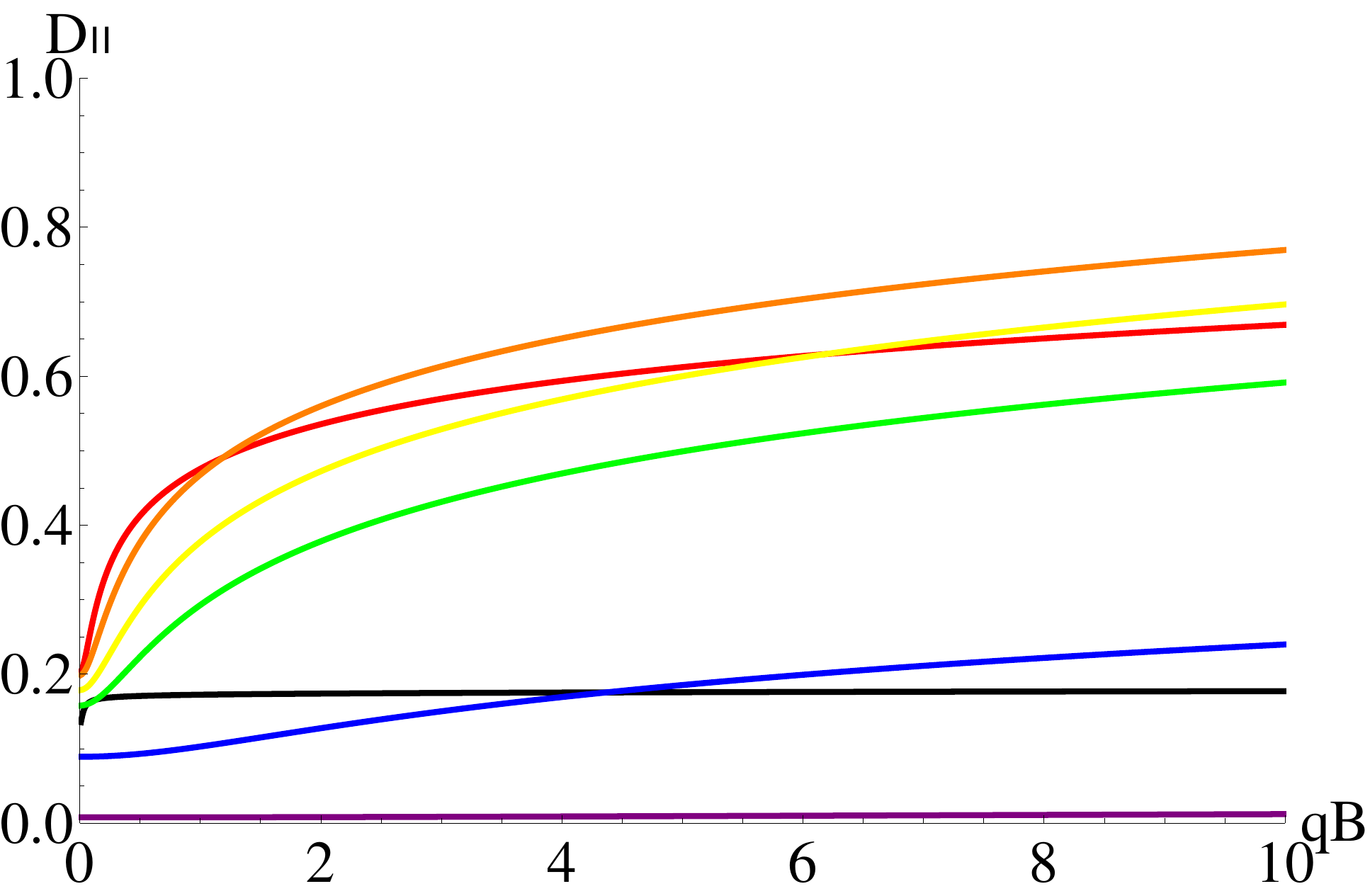}
\caption{Heavy quark diffusion coefficient for propagation parallel to $\mathbf{B}$ as a function of applied magnetic field $qB$. Black: $t=0.13$, red: $t=0.25$, orange: $t=0.35$, yellow: $t=0.45$, green: $t=0.55$, blue: $t=1.1$, purple: $t=2.2$.}
\label{diff3Tb}
\end{minipage}
\hfill
\begin{minipage}{0.48\textwidth}
\centering
\includegraphics[width=0.95\linewidth]{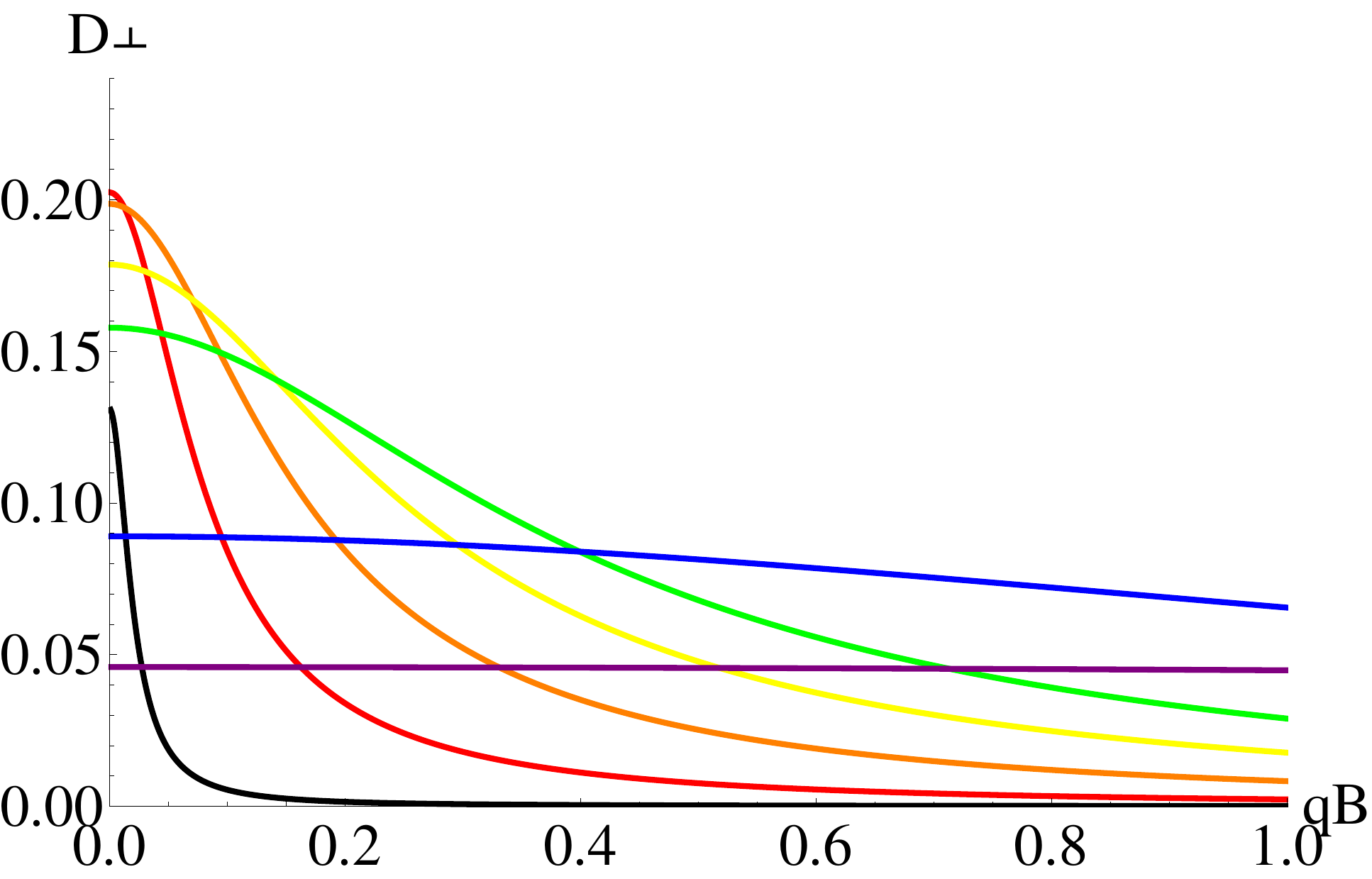}
\caption{Heavy quark diffusion coefficient for propagation transverse to $\mathbf{B}$ as a function of applied magnetic field $qB$. Black: $t=0.13$, red: $t=0.25$, orange: $t=0.35$, yellow: $t=0.45$, green: $t=0.55$, blue: $t=1.1$, purple: $t=2.2$.}
\label{diff1Tb}
\end{minipage}
\end{figure}

\begin{figure}[h]
\begin{minipage}{0.48\textwidth}
\centering
\includegraphics[width=0.95\linewidth]{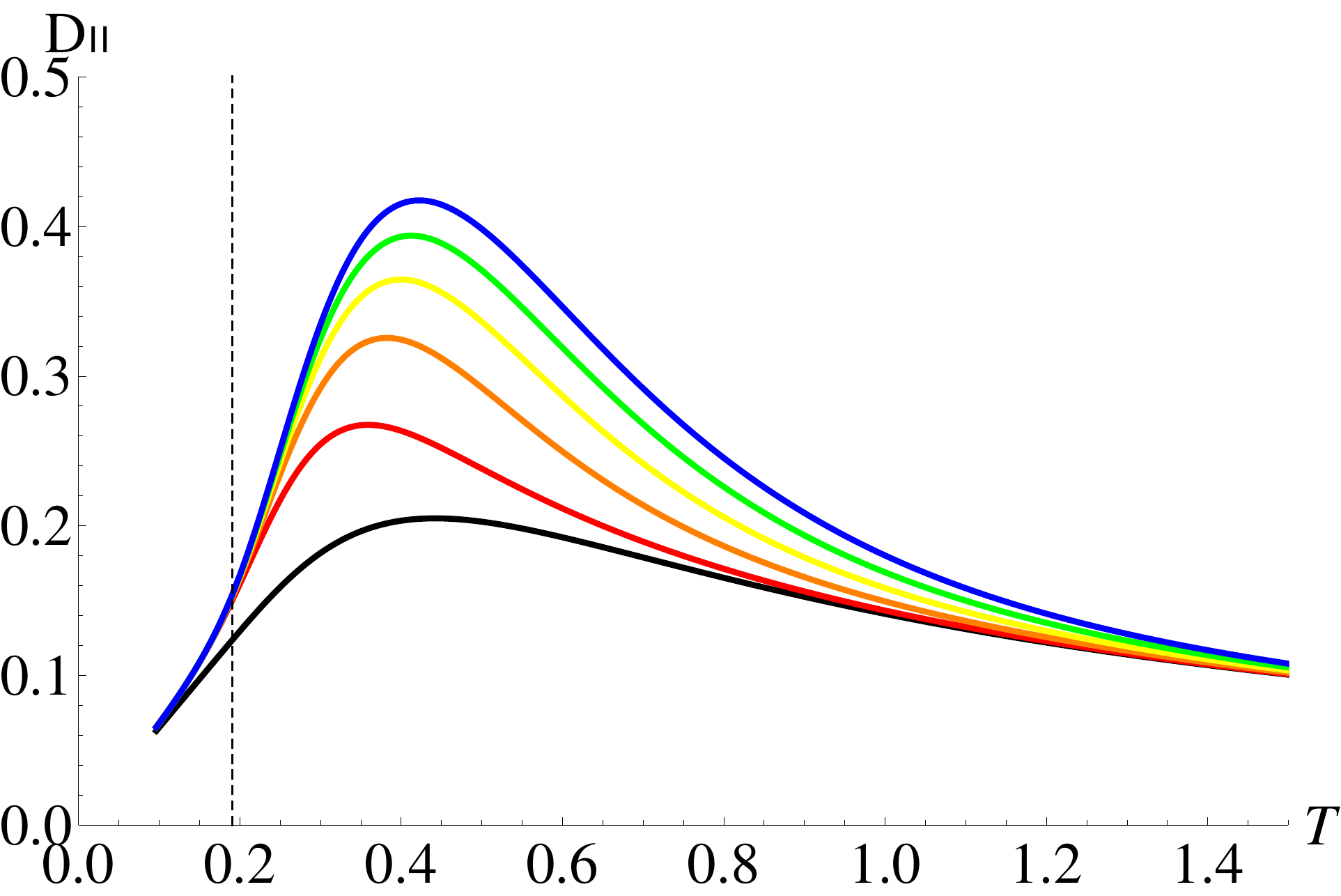}
\caption{Heavy quark diffusion coefficient for propagation parallel to $\mathbf{B}$ as a function of temperature $T$. Black: $qB=0$ GeV$^2$, red: $qB= 0.1$ GeV$^2$, orange: $qB= 0.2$ GeV$^2$, yellow: $qB= 0.3$ GeV$^2$, green: $qB= 0.4$ GeV$^2$, blue: $qB= 0.5$ GeV$^2$.}
\label{diff3Tc}
\end{minipage}
\hfill
\begin{minipage}{0.48\textwidth}
\centering
\includegraphics[width=0.95\linewidth]{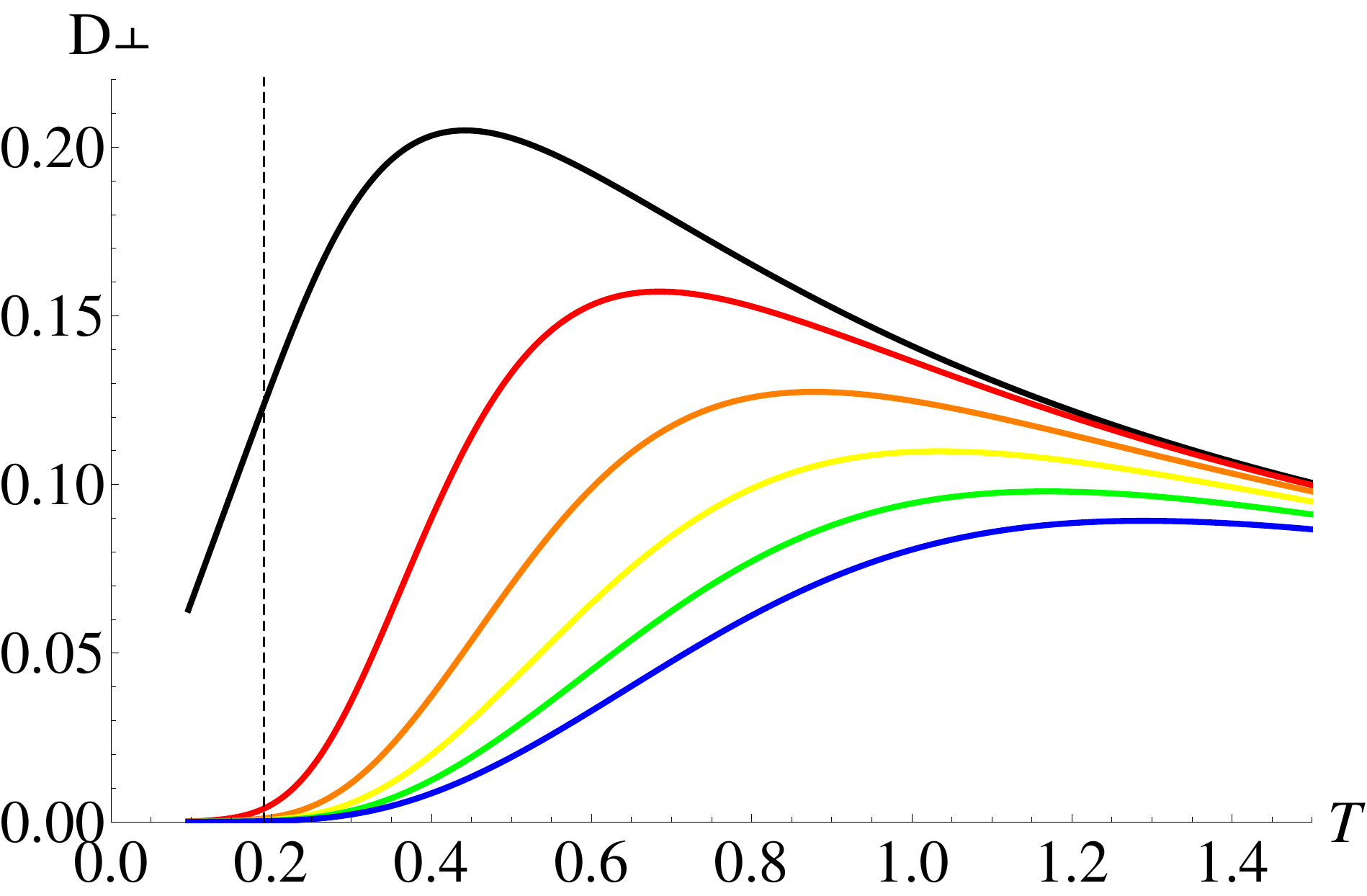}
\caption{Heavy quark diffusion coefficient for propagation transverse to $\mathbf{B}$ as a function of temperature $T$. Black: $qB=0$ GeV$^2$, red: $qB= 0.1$ GeV$^2$, orange: $qB= 0.2$ GeV$^2$, yellow: $qB= 0.3$ GeV$^2$, green: $qB= 0.4$ GeV$^2$, blue: $qB= 0.5$ GeV$^2$.}
\label{diff1Tc}
\end{minipage}
\end{figure}

The reason for this behavior can be understood even at $qB=0$ and is an inherent feature of the soft wall model. At $qB=0$, the diffusion coefficient can be easily computed analytically:
\begin{equation}
\label{anab0}
D = D_{\perp} = D_{\parallel} = \frac{1-e^{-cz_h^2}}{2 z_h c}.
\end{equation}
For $t$ small, we have $D\sim\frac{T}{c}$, whereas for $T$ large, one finds $D\sim\frac{1}{T}$. The turning point of this behaviour is around $t\approx 0.3$, about three times the deconfinement temperature. $\mathcal{N}=4$ SYM has $D\sim\frac{1}{T}$ throughout the entire temperature range, due to conformal invariance. In the high temperature regime, all mass scales are negligible, and it is natural we agree with the SYM story. Even more so since we are working in the quenched approximation and the glue sectors of $\mathcal{N}=4$ SYM and QCD are identical.

It is necessary that $D\to 0$ as $T\to0$ since the black hole then disappears, and we are reduced to the confining phase at $T=0$. Since the temperature does not figure in any of the holographic 2-point correlators in the thermal gas phase, this is the result of the confining phase at any $T$. And since we also expect, from the QCD viewpoint, that $D=0$ at low temperatures since no free quarks exist anymore in the confining phase, we can foresee the $T\to0$ limit in the black hole phase to yield $D=0$, a property that is indeed borne out when including the soft wall. In the SYM case ($c\to 0$), the theory is in the deconfining phase for any $T$, and a similar criterion does not exist. \\
It was explained in \cite{deBoer:2008gu} that a behavior of $D\to +\infty$ as $T\to 0$ actually comes from the absence of a diffusive regime; the particle only experiences the ballistic regime for which effectively $D \to + \infty$. This works fine as long as no confinement sets in. If it does, then this regime is inaccessible as the heavy quark will get bound in a color invariant state before moving in the thermal medium. \\
Note that in a magnetic field, the above behavior of $\mathcal{N}=4$ SYM changes dramatically: $D\to0$ now as $T\to 0$, check \eqref{check} in the next Section. This can be understood from physical intuition as well, as the magnetic field forces the charged particle to follow a helical trajectory, preventing its ballistic escape, and causing $D$ to effectively vanish in this case. Again in a more realistic QCD scenario, also this feature should be absent as the colored particle will still get confined in hadrons instead.
We conclude that the soft wall is hence crucial for obtaining a behavior in tune with expectations from a realistic thermal theory with (de)confinement features, such as QCD.

One further notices that at high $T$, the curves flatten out. High temperature washes out the influence of the magnetic field, as expected. This is also corroborated from a Langevin analysis, where one expects the friction coefficient to scale as $\gamma \sim T^2$, and hence the $B$-dependent contribution becomes negligible for large $T$ as it becomes apparent from \eqref{lang1}. The high-temperature limit has universal behavior:
\begin{equation}
\chi = \frac{N_c}{3}T^2, \quad D_\parallel = D_\perp = \frac{1}{2\pi T}.
\end{equation}

For small $T$, the relevant quantities behave as
\begin{alignat}{3}
\chi &\sim ce^{-cz_h^2}, \quad &&D \sim \frac{T}{c} , \quad &qB = 0 ,\\
\chi &\sim cz_h^2 e^{-cz_h^2} qB , \quad &&D_{\parallel}  \sim \frac{T}{c}, \quad D_{\perp} \sim \frac{T^5}{c (qB)^2}, \quad &qB \neq 0,
\end{alignat}
as can be seen in Figures~\ref{diff3Tc} and \ref{diff1Tc}.

\begin{figure}[h]
\begin{minipage}{0.48\textwidth}
\centering
\includegraphics[width=0.95\linewidth]{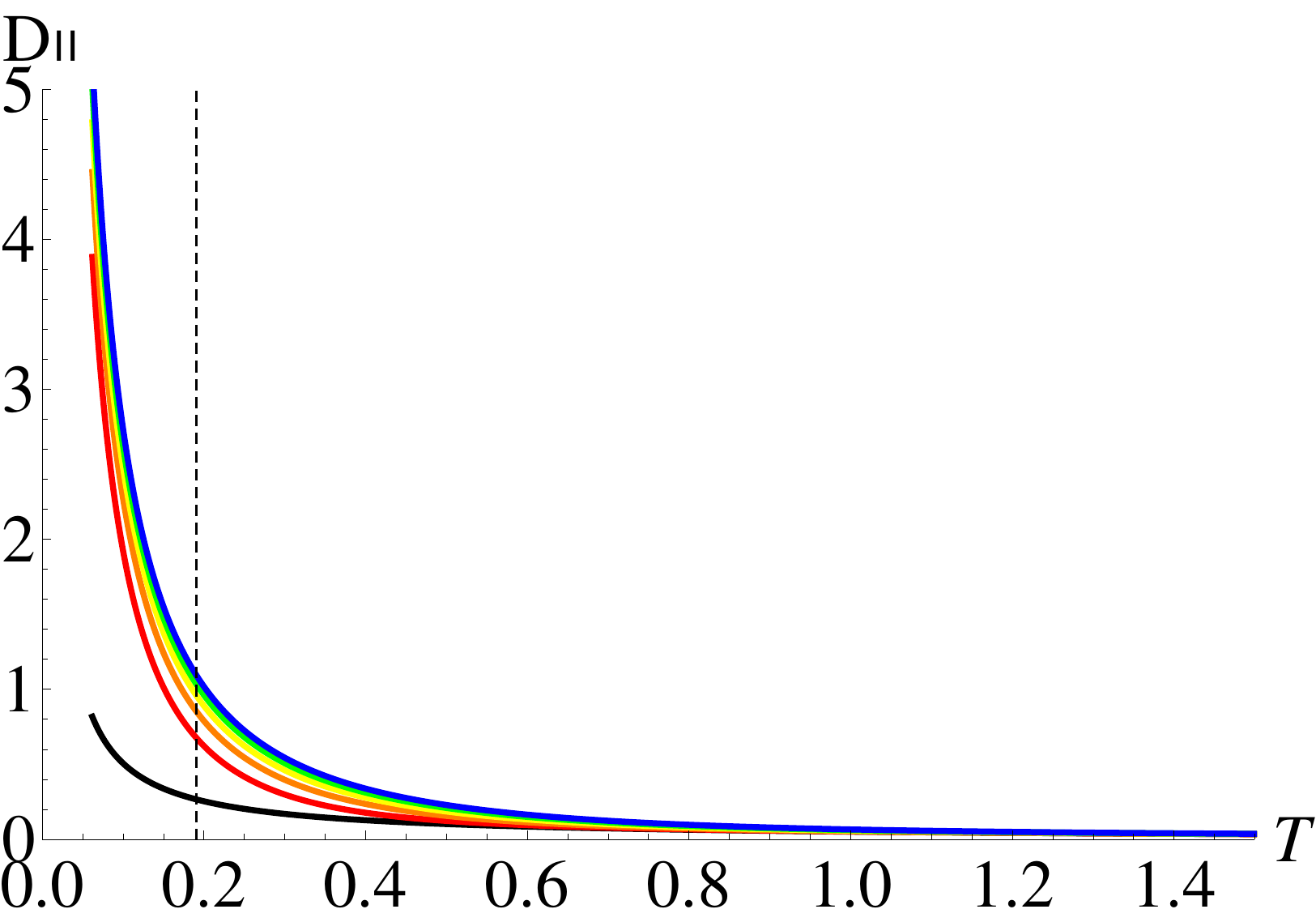}
\caption{Heavy quark diffusion coefficient for $c=0$ for propagation parallel to $\mathbf{B}$ as a function of temperature $T$. Black: $qB=0$ GeV$^2$, red: $qB= 0.1$ GeV$^2$, orange: $qB= 0.2$ GeV$^2$, yellow: $qB= 0.3$ GeV$^2$, green: $qB= 0.4$ GeV$^2$, blue: $qB= 0.5$ GeV$^2$.}
\label{diff3Tc0}
\end{minipage}
\hfill
\begin{minipage}{0.48\textwidth}
\centering
\includegraphics[width=0.95\linewidth]{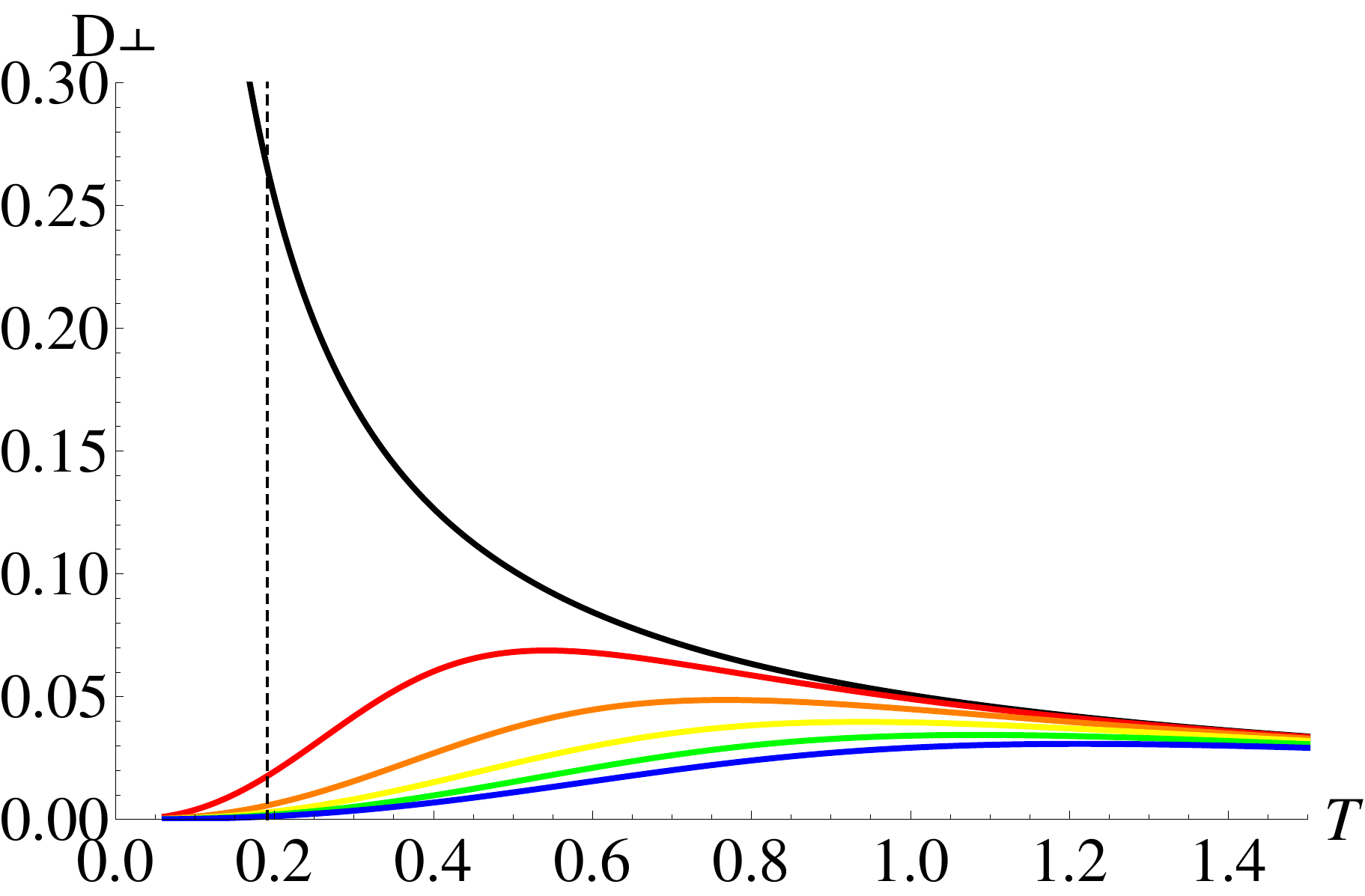}
\caption{Heavy quark diffusion coefficient for $c=0$ for propagation transverse to $\mathbf{B}$ as a function of temperature $T$. Black: $qB=0$ GeV$^2$, red: $qB= 0.1$ GeV$^2$, orange: $qB= 0.2$ GeV$^2$, yellow: $qB= 0.3$ GeV$^2$, green: $qB= 0.4$ GeV$^2$, blue: $qB= 0.5$ GeV$^2$.}
\label{diff1Tc0}
\end{minipage}
\end{figure}

For later comparison with a stringy calculation of the diffusion, we can also formally take the $c\to0$ limit, with results shown in Figures~\ref{diff3Tc0} and \ref{diff1Tc0}.
On the other hand, the heavy quark mass itself is modeled in through $c$, so taking the limit $c\to 0$ is perhaps not so instructive in the end.

For non-zero $B$, it is yet impossible to compare to lattice estimates as these are at the moment nonexistent. Though, for $B=0$, we may make contact with the results of \cite{Banerjee:2011ra,Francis:2015daa}.  What is apparent is that, for large $T$, the $D\sim 1/T$ behavior is violated at least just above $T_c$ and it seems more likely that a slower decrease (or even an increase) with $T$ is present. As computing the diffusion constants from a numerical lattice simulation is a highly complicated task, future improved results might further pin down the temperature dependence of the diffusion constants. General remarks concerning this can be found in \cite{Petreczky:2005nh}. There is general consensus that at least $D$ should be not much larger than $1/T$ in order to accommodate for the measured elliptic flow parameter. For the record, for $B=0$, our results are compatible with an increasing $D$ around $T_c$, as it is apparent from Figures~\ref{diff3Tc} or \ref{diff1Tc}, but given that $T_c=0.191~\text{GeV}$, $D$ is certainly not much larger than $1/T_c$ around $T_c$.

\section{Heavy quark diffusion from the hanging string approach}\label{sect6}

One of the first holographic studies of heavy quark diffusion is \cite{CasalderreySolana:2006rq}. A holographic study, using the probe string attached to both the boundary and the horizon, can be found in \cite{deBoer:2008gu,Fischler:2012ff}. This study is however restricted to the (better controlled) AdS/CFT case ($\mathcal{N}=4$ SYM on the boundary) corresponding to letting $c\to0$ in the soft wall model. In \cite{Fischler:2012ff}, one obtains as the result for $D_\perp$:\footnote{$\lambda$ is the 't Hooft parameter in $\mathcal{N}=4$ SYM.}
\begin{equation}\label{check}
D_\perp = \frac{2\pi \sqrt{\lambda}T^3 }{4B^2 + \pi^2 \lambda T^4}, \quad \sqrt{\lambda} = \frac{L^2}{\alpha'}
\end{equation}
which is qualitatively of the same shape as we found before using our spectral method, at least when $qB$ is not too small. The procedure is in some sense more in the particle language (first quantization) as the heavy quark is represented explicitly as the endpoint of a dangling string. We on the other hand modeled the heavy quark in second quantization using the standard holographic dictionary in the first part of this paper.

\begin{figure}[h]
\centering
\includegraphics[width=0.4\linewidth]{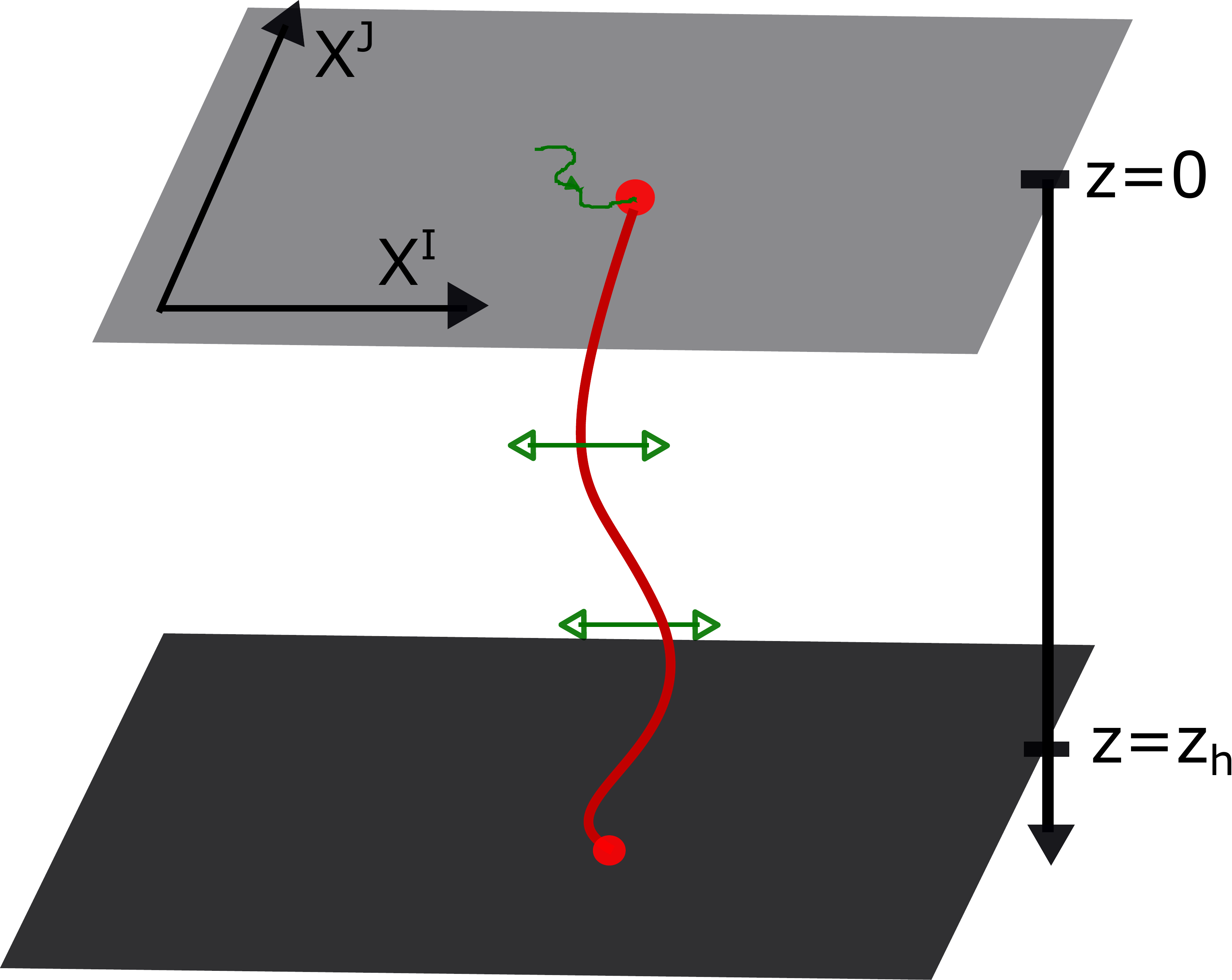}
\caption{Cartoon of the situation: a long string is suspended from the boundary ($z=0$) to the horizon ($z=z_h$). The Hawking--Unruh effect creates transverse wiggles along the string that get transported back to the boundary where the endpoint experiences Brownian motion as a result \cite{deBoer:2008gu}.}
\label{scheme}
\end{figure}

Let us have a closer look at this method that envisions a heavy quark in a thermal medium as being holographically dual to a string suspended from the boundary into the black hole horizon (Figure~\ref{scheme}).
This classical string gets its dynamics from the Nambu-Goto action. This action contains only the string frame metric however, and no coupling to the dilaton is present. Hence the soft wall has no influence whatsoever on the resulting Brownian motion of the heavy quark and consequently the diffusion coefficient is the same as in $\mathcal{N}=4$ SYM. This is of course unphysical, and this is due to the fact that the soft wall model ignores backreaction of the wall on the metric background.

As a first remedy to this (besides of course constructing a backreacted soft wall model), one is tempted to just insert the dilaton wall into the Nambu-Goto action and see what happens. From a top-down approach this is not correct, but from a bottom-up approach, this makes the string worldsheet action obtain the same amount of damping as the bulk field actions would experience.\footnote{The situation can be a bit compared with the conjectured introduction of the dilaton, including that of the soft wall model, in the Ryu-Takayanagi minimal area prescription for holographic entanglement entropy, as proposed in \cite{Klebanov:2007ws}.}
This is similar to what we did with the DBI-model itself: ordinarily this is constructed as the effective action describing the embedding and gauge fields propagating on the world-volume of a D-brane. The relevant metric in this action is again the string frame metric. But we constructed our model by including the soft wall directly in this action. Since we obtained good results with this ansatz, it seems at least worthwhile to check what happens in the string case. An alternative approach would be to use the alternative AdS/QCD model considered in \cite{Andreev:2006ct,Andreev:2006nw} and related works, in which an exponential warp factor is present in the metric itself (and only there) with the opposite sign for the dilaton, a feature leading to an area law for the Wilson loop, in contrast with the original wall model. Though, this ``other sign dilaton'' has also been met with criticism, see \cite{Karch:2010eg}. Notice also that this ``other sign dilaton'' corresponds to a growing exponential in the action when probing deeper into the AdS bulk. We will not further consider the latter model in the current paper.

\subsection{Without magnetic field}
We will closely follow the analysis of \cite{Fischler:2012ff}. In general, considering a one-dimensional motion in direction $X^I$ governed by a Langevin equation allows to extract the diffusion constant $D$, see e.g.~\cite{kt}, from the mean displacement squared. More precisely,
\begin{eqnarray}\label{displ}
D= \lim_{t\to\infty} \frac{\braket{ (X^I(t)-X^I(0))^2}}{2t}.
\end{eqnarray}
We will effectively compute the quantity $\braket{ (X^I(t)-X^I(0))^2}$ where $X^I(t)$ corresponds to the end point of the hanging string. The latter is described by a Nambu-Goto action, now augmented with the soft wall term as motivated before:
\begin{align}
S &= - \frac{1}{2\pi \alpha'}\int d^2 \sigma e^{-\Phi}\sqrt{-\det \gamma_{\mu\nu}} \\
&\approx  - \frac{1}{4\pi \alpha'}\int d^2 \sigma \sqrt{-g}e^{-\Phi} g^{\mu\nu}G_{IJ} \frac{\partial X^I}{\partial x^\mu}\frac{\partial X^J}{\partial x^\nu},
\end{align}
where $g_{\mu\nu}$ is the metric in the $(t,z)$ plane and $G_{IJ}$ the metric in the remaining transverse directions. We expanded this action for small $\partial_t X^I(t)$ up to the quadratic term. The constant term in this action was dropped, and corresponds to the mass of the string as (imposing a cut-off at $z=z_c \ll 1$):\footnote{The $g_{00}$ weighing of the integral represents the energy redshift at different radial locations.}
\begin{equation}
m = \frac{1}{2\pi \alpha'} \int_{z_c}^{z_h}dz e^{-\Phi}\sqrt{g_{00}g_{zz}} = \frac{1}{2\pi \alpha'} \int_{z_c}^{z_h}dz \frac{L^2}{z^2}e^{-cz^2} \approx \frac{L^2}{2\pi\alpha' z_c}.
\end{equation}
The equations of motion for the transverse coordinates $X^I$ are given by
\begin{equation}
\partial_\mu\left(g^{\mu\nu}\sqrt{-g}e^{-cz^2}G_{IJ}\partial_\nu X^J\right) = 0, \quad I=1 \ldots 3.
\end{equation}
Next we determine the explicit solutions to this equation in the low frequency approximation (as this will be all we need). The following analysis is a bit technical. As a guideline, the result is that the quantum field (\ref{mode}), containing parameters $\bar{A}$ (\ref{abar}) and $\bar{B}$ (\ref{bbar}), is used in the 2-point correlator (\ref{conthere}). The reader who is not interested in the technical details can skim the next paragraphs and continue reading from (\ref{conthere}).

\subsubsection{Mode expansion of the quantum field}

Now the detailed analysis. In our case, we have
\begin{align}
g_{00} = -\frac{L^2}{z^2}\left(1-\frac{z^4}{z_h^4}\right), \quad g_{zz} = \frac{L^2}{z^2\left(1-\frac{z^4}{z_h^4}\right)}, \quad \sqrt{-g} = \frac{L^2}{z^2}, \quad G_{IJ} = \delta_{IJ}\frac{L^2}{z^2}.
\end{align}
For $\partial_0 = -i\omega$, one finds explicitly
\begin{equation}
\frac{L^2}{z^2\left(1-\frac{z^4}{z_h^4}\right)}e^{-cz^2}\omega^2 X^I + \partial_z\left(\frac{L^2}{z^2}\left(1-\frac{z^4}{z_h^4}\right)e^{-cz^2}\partial_z X^I\right) = 0.
\end{equation}
Setting $\rho= \frac{z}{z_h}$ and $\nu = \omega z_h$, this equation for a general direction $\phi = X^I$ is rewritten in dimensionless quantities as
\begin{equation}
\label{danglode}
\frac{\nu^2}{\rho^2\left(1-\rho^4\right)}e^{-cz_h^2\rho^2} \phi + \partial_\rho\left(\frac{1}{\rho^2}\left(1-\rho^4\right)e^{-ccz_h^2\rho^2}\partial_\rho \phi\right) = 0.
\end{equation}
The horizon is at $\rho=1$ and the boundary at $\rho=0$ in these coordinates.

Again resorting to a low-frequency (hydrodynamic) expansion, one writes: 
\begin{equation}
\label{hydroexp}
\phi(t,\rho) = \phi_0(t,\rho) + \nu \phi_1(t,\rho) + \mathcal{O}(\nu^2),
\end{equation}
where the solution can be readily written down as
\begin{align}
\phi_0(t,\rho) &= B_1 + B_2 \int_{0}^{\rho} du \frac{u^2}{1-u^4}e^{cz_h^2u^2}, \\
\phi_1(t,\rho) &= B_3 + B_4 \int_{0}^{\rho} du \frac{u^2}{1-u^4}e^{cz_h^2u^2},
\end{align}
for arbitrary integration constants $B_1$, $B_2$, $B_3$ and $B_4$.
Near the horizon ($\rho \approx 1$), one finds the asymptotic behavior for solutions of (\ref{danglode}) to be:
\begin{equation}
\phi \sim (1-\rho)^{\pm\frac{i\nu}{4}}.
\end{equation}
The outgoing mode has the plus sign and the ingoing mode the minus sign:
\begin{align}
g^{\text{out}} &= e^{i\frac{\nu}{4}\ln(1-\rho)}e^{-i\omega t}, \\
g^{\text{in}} &= e^{-i\frac{\nu}{4}\ln(1-\rho)}e^{-i\omega t}.
\end{align}
Expanding the hydrodynamic expansion (\ref{hydroexp}) near the horizon, and matching these expansions, then leads to
\begin{align}
g^{\text{out/in}} = 1 \pm i\nu e^{-cz_h^2}\left(C - \int_{0}^{\rho}du \frac{u^2}{1-u^4}e^{cz_h^2u^2}\right) + \mathcal{O}(\nu^2)
\end{align}
as the low-frequency approximation to the mode that is ingoing/outgoing at the horizon, but now considered throughout the entire space. \\
Here, we wrote $C = \int_{0}^{1}du \left(\frac{u^2e^{cz_h^2u^2}}{1-u^4} - \frac{e^{cz_h^2}}{4(1-u)}\right)$ as the finite term of the above integral in the $\rho \to 1$ expansion.
The final step is to match this to the asymptotic near-boundary modes.
Expanding these functions for $\rho\approx 0$, one obtains
\begin{align}
g^{\text{out/in}} \approx \left(1 \pm i\nu e^{-cz_h^2}\left(C - \frac{1}{3}\rho^3 - \frac{1}{5}cz_h^2\rho^5\right) + \mathcal{O}(\rho^7)\right)+ \mathcal{O}(\nu^2).
\end{align}
Near the boundary ($\rho\approx 0$), one instead finds from a Frobenius analysis of (\ref{danglode}):
\begin{equation}
\phi \sim \rho^0 \quad \text{or} \quad  \phi \sim \rho^3,
\end{equation}
leading to the series solutions\footnote{The coefficient of the logarithmic series happens to vanish.}
\begin{align}
\psi_1 &= \rho^3 + \frac{6c-\nu^2}{10}\rho^5 + \left(\frac{3}{7}+\frac{3}{14}c^2z_h^4-\frac{2}{35}cz_h^2\nu^2+\frac{1}{280}\nu^4\right)\rho^7 + \mathcal{O}(\rho^9), \\
\psi_2 &= 1 + \frac{\nu^2}{2}\rho^2 + \left(\frac{cz_h^2 \nu^2}{2} - \frac{\nu^4}{8}\right)\rho^4 + \mathcal{O}(\rho^6).
\end{align}
The ingoing and outgoing solutions then match to these solutions as
\begin{equation}
g^{\text{out/in}} = C_1 \psi_1 + C_2 \psi_2,
\end{equation}
with $C_1 = \mp \frac{i\nu e^{-cz_h^2}}{3} + \mathcal{O}(\nu^2)$ and $C_2 = 1 \pm i \nu  e^{-cz_h^2} C + \mathcal{O}(\nu^2)$.

The only features of this solution we will need are the near-boundary expansion of the low-frequency asymptotics:
\begin{alignat}{2}
g^{\text{out/in}} &\approx 1 , \quad &&\rho \approx 0, \\
\partial_\rho g^{\text{out/in}} &\approx \mp i\nu e^{-cz_h^2}\rho^2 , \quad &&\rho \approx 0.
\end{alignat}
Note that the low-frequency limit is taken \emph{before} the $\rho \to 0$ limit; more on this below.

The mode expansion of the quantum field is given by
\begin{equation}\label{mode}
X^I(t,z) = \sum_{\omega>0} \left[a_\omega f_\omega(t,z) + a_\omega^{\dagger}f_\omega^*(t,z)\right],
\end{equation}
where each mode is given by
\begin{equation}
f_\omega(t,z) = \bar A\left[g^{\text{out}}(z) + \bar B g^{\text{in}}(z)\right]e^{-i\omega t}.
\end{equation}
From this, we can identify the position of the heavy quark with the endpoint of the string, viz.~
\begin{equation}
X^I(t)=X^I(t,\rho_c)
\end{equation}
upon switching to the $\rho$ coordinate.

The factor $\bar B$ can be determined by imposing Neumann boundary conditions at the holographic boundary if no boundary magnetic field is present. Let us first sketch the derivation in this case.

It is important to clarify what the order of the limits is. If one takes $\rho_c \to 0$ first, then this leads to $\bar B$ being a phase:
\begin{equation}
\bar B = -\frac{\partial_\rho g^{\text{out}}(t,0)}{\partial_\rho g^{\text{in}}(t,0)} = -\frac{1+i\nu C e^{-cz_h^2}}{1-i\nu C e^{-cz_h^2}}.
\end{equation}
However, $\rho_c$ has to be small but non-zero in order to have a finite-mass heavy quark. So other limits (such as the small frequency limit) have to be taken first. With this knowledge, $\bar B$ is instead given by
\begin{equation}
\label{bbar}
\bar B = -\frac{\partial_\rho g^{\text{out}}(t,\rho_c)}{\partial_\rho g^{\text{in}}(t,\rho_c)} = -\frac{i\nu e^{-cz_h^2}\rho_c^2}{-i\nu e^{-cz_h^2}\rho_c^2} + \mathcal{O}(\nu) = 1 + \mathcal{O}(\nu).
\end{equation}

Additionally imposing Neumann boundary conditions at $\rho = 1-\epsilon$, close to the horizon, discretizes the spectrum, and one finds $\bar B = \epsilon^{i\nu/2}$. This then, in turn, leads to a discretization of the spectrum with:
\begin{equation}
\Delta \nu = \frac{4\pi}{\ln(1/\epsilon)},
\end{equation}
or, by restoring the units (rescaling by $z_h$),
\begin{equation}
\Delta \omega = \frac{4\pi^2}{\beta\ln(1/\epsilon)}\,,\qquad \beta=\frac{1}{T}.
\end{equation}
As a consequence, we have
\begin{equation}
 \sum_{\omega>0} \, \to \, \ln(1/\epsilon)\frac{\beta}{4\pi^2}\int_0^\infty d\omega\qquad\text{for}~\epsilon\to0.
\end{equation}
The prefactor $\bar A$ can now be determined by imposing canonical normalization of the fields. The Klein-Gordon inner product in this case, evaluated on a constant $t$-slice is given by: 
\begin{equation}\label{KG}
(f,g) = - \frac{i}{2\pi\alpha'}\int dz \sqrt{-g}e^{-\Phi}\left|g^{00}\right|\frac{L^2}{z^2}\left(f \partial_0 g^{*} - \partial_0 f g^{*}\right).
\end{equation}
If we normalize the mode functions such that $(f_\omega, f_\omega) = 1$, then the canonical quantization yields
\begin{equation}\label{comm}
\left[\phi(t,\mathbf{x}), \pi(t,\mathbf{y})\right] = i\delta (\mathbf{x}-\mathbf{y}) \quad \Longleftrightarrow \quad \left[a_\omega, a_{\omega'}^\dagger\right] = \delta_{\omega,\omega'}.
\end{equation}
In our case, the constant $\bar A$ is then found as
\begin{equation}
\label{abar}
\frac{\omega}{2\pi \alpha'}\frac{L^2}{z_h}\ln(1/\epsilon)\left|\bar A\right|^2 e^{-cz_h^2} =1.
\end{equation}
To obtain this result, we took into account that the main contribution to the integral \eqref{KG} arises from the logarithmic (infrared) near-horizon divergence, regularized by $\epsilon$. The UV singularity from $z=0$ is cut-off in a natural way by $z_c$, which is directly related to the heavy quark mass. Although we will not consider this limit here, the static limit (corresponding to an infinitely heavy quark) should from this perspective only be taken at the very end, to maintain the dominance of the near-horizon. This is similar to the previously discussed order of limits.

\subsubsection{Diffusion coefficient}
The standard commutator relation given in \eqref{comm}, together with the mode expansion \eqref{mode} in the $\epsilon\to0$ limit, leads after some algebra to
\begin{equation}
\label{conthere}
\Braket{:X^I(t) X^I(0):}=\frac{\beta}{4\pi^2}\int_0^\infty \frac{d\omega}{\omega} \frac{2|\bar  A'|^2\cos (\omega t)}{e^{\beta \omega}-1}\left|g^{\text{out}}(\rho_c) + \bar B g^{\text{in}}(\rho_c)\right|^2
\end{equation}
for the (normal-ordered) propagator, with $|\bar A'|^2= \ln(1/\epsilon)|\omega\bar A|^2$. The quantum state (or better said, density matrix) to be used here is the thermal ensemble, as follows by standard Hawking--Unruh arguments (see e.g. \cite{deBoer:2008gu} for more details in this context). From this equation, the mean displacement squared follows as
\begin{eqnarray}
\braket{:(X^I(t)-X^I(0))^2:}=\frac{2\beta}{\pi^2}\int_0^\infty \frac{d\omega}{\omega} \frac{|\bar  A'|^2\sin^2(\omega t/2)}{e^{\beta \omega}-1}\left|g^{\text{out}}(\rho_c) + \bar B g^{\text{in}}(\rho_c)\right|^2.
\end{eqnarray}
As the integral for $t\to \infty$ is dominated by small $\omega$, one finds that
\begin{eqnarray}
\int_0^\infty \frac{d\omega}{\omega} \frac{\sin^2(\omega t/2)}{e^{\beta \omega}-1}\to \frac{\pi}{4}\frac{t}{\beta}\qquad\text{for}~t\to\infty,
\end{eqnarray}
so that, essentially, up to a prefactor, the diffusion constant $D$ is determined by the value of $\left|g^{\text{out}}(\rho_c) + \bar B g^{\text{in}}(\rho_c)\right|^2$. This was precisely the reason it was sufficient to only consider the low frequency approximation to the mode solutions.

In the low-frequency limit, one obtains
\begin{equation}
\label{modsq}
\left|g^{\text{out}} + \bar{B} g^{\text{in}}\right|^2 \approx 4 + \mathcal{O}(\nu)
\end{equation}
since $g^{\text{in/out}} \approx 1$ and $\bar{B} \approx 1$, the only appearance of the soft wall in this entire computation, would come from $\bar A'$, and would hence only multiply the final correlator with $e^{cz_h^2}$. The diffusion coefficient is hence given by
\begin{equation}
D = D_{c=0}e^{cz_h^2} = \frac{2\alpha'}{\pi L^2 T}e^{\frac{c}{\pi^2 T^2}},
\end{equation}
and is hence larger than when $c=0$. An intuitive explanation is as follows. Close to the horizon, the Nambu-Goto action attains an effective string tension $\tau_{\text{eff}} \sim \tau e^{-cz_h^2}$. As the string tension decreases, the string becomes floppier, and fluctuations have larger amplitudes. Hence diffusion proceeds easier.

It is to be remarked that at high temperatures ($z_h \to 0$), the hanging string and the spectral function approach (discussed in previous Sections) obviously agree.
At low temperature however, the results disagree, where we previously found that $D \sim T/c$ at low temperatures. This can be understood since the hanging string picture requires us to treat the heavy quark in a first-quantized fashion and for very large mass. The spectral function approach on the other hand seems better suited for quark masses that are not arbitrarily large.

\subsection{With magnetic field}
Next, we introduce the magnetic field. The Neumann boundary condition at $\rho=0$ then gets modified into mixed boundary conditions, see \cite{Fischler:2012ff},
\begin{equation}
\sqrt{-g_{00}}\sqrt{g^{zz}}G_{IJ}\partial_z X^J = 2\pi \alpha' F_{IJ}\partial_t X^J.
\end{equation}
This corresponds to a Lorentz force acting on the (charged) end point of the string.

Also here, we will put the magnetic field along the spatial 3-direction. As there is no influence of such a field on the motion along this direction, we only need to consider diffusion in the remaining transverse $(1,2)$ directions. Note that this is already a difference with our previous spectral function approach, where we found a non-trivial influence on both $D_\parallel$ and $D_\perp$.
One finds here:\footnote{In fact, to be precise we would have to replace here $B \to \frac{2}{3}q B$ to include the charge of the charm quark, but we will not do so for the sake of notational convenience.}
\begin{align}
\left.\left(1-\frac{z^4}{z_h^4}\right)\frac{L^2}{z^2}\partial_z X_1\right|_{z=\epsilon} &= \left.2\pi\alpha' B (-i\omega) X_2\right|_{z=\epsilon}, \\
\left.\left(1-\frac{z^4}{z_h^4}\right)\frac{L^2}{z^2}\partial_z X_2\right|_{z=\epsilon} &= \left.-2\pi\alpha' B (-i\omega) X_1\right|_{z=\epsilon}.
\end{align}
which can be diagonalized using $X_\pm = X_1 \pm i X_2$:
\begin{align}
\left.\left(1-\frac{z^4}{z_h^4}\right)\frac{L^2}{z^2}\partial_z X_+\right|_{z=\epsilon} &= \left.-2\pi\alpha' B \omega X_+ \right|_{z=\epsilon}, \\
\left.\left(1-\frac{z^4}{z_h^4}\right)\frac{L^2}{z^2}\partial_z X_-\right|_{z=\epsilon} &= \left.2\pi\alpha' B \omega X_- \right|_{z=\epsilon}.
\end{align}
Both $X_\pm$ then satisfy the above mode decompositions, and can be treated as elementary quantum fields.

A completely similar analysis as before applies now. Using the near-boundary asymptotics, one finds that (\ref{modsq}) gets modified in the low-frequency regime:\footnote{For two different coefficients $\bar B_\pm$ associated to $X_\pm$ \cite{Fischler:2012ff}.}
\begin{equation}
\left|1+\bar B_\pm\right|^2 = \frac{4L^4\pi^2 T^4e^{-2cz_h^2}}{L^4\pi^2 T^4e^{-2cz_h^2} + 4\alpha'^2B^2}.
\end{equation}
Hence, in the end, the (transverse) diffusion coefficient one finds will be of the form:\footnote{The $c\to 0$ limit agrees with (\ref{check}) as it should.}
\begin{equation}
\label{Dhang}
D_\perp = 2\pi \frac{\alpha'L^2T^3 e^{-cz_h^2}}{L^4 \pi^2 T^4e^{-2cz_h^2} + 4\alpha'^2B^2},
\end{equation}
implying in this case, that as $T \to 0$, $D_\perp\to 0$ as well, unless $B=0$ in which case $D_\perp$ diverges again at zero temperature. Note that the low-temperature region of this equation is again an extrapolation as one should actually resort to the confined geometry in this case. This remark also holds for the Figure we present below. \\
To compare this result with the spectral function result \eqref{Dperpe}, we need to fix $\alpha'/L^2$. A priori, we are not guaranteed that this ratio is the same as before, as we took an entirely different starting point. For $B=0$ and at high temperature, we find that \eqref{Dhang} reduces to $D_\perp \to \frac{2\alpha'}{L^2 T}$, agreeing with \eqref{anab0} as $D_\perp \to 1/(2\pi T)$ provided $\alpha'/L^2=1/4$, and hence this is the value we choose here.

\begin{figure}[h]
\centering
\includegraphics[width=0.6\linewidth]{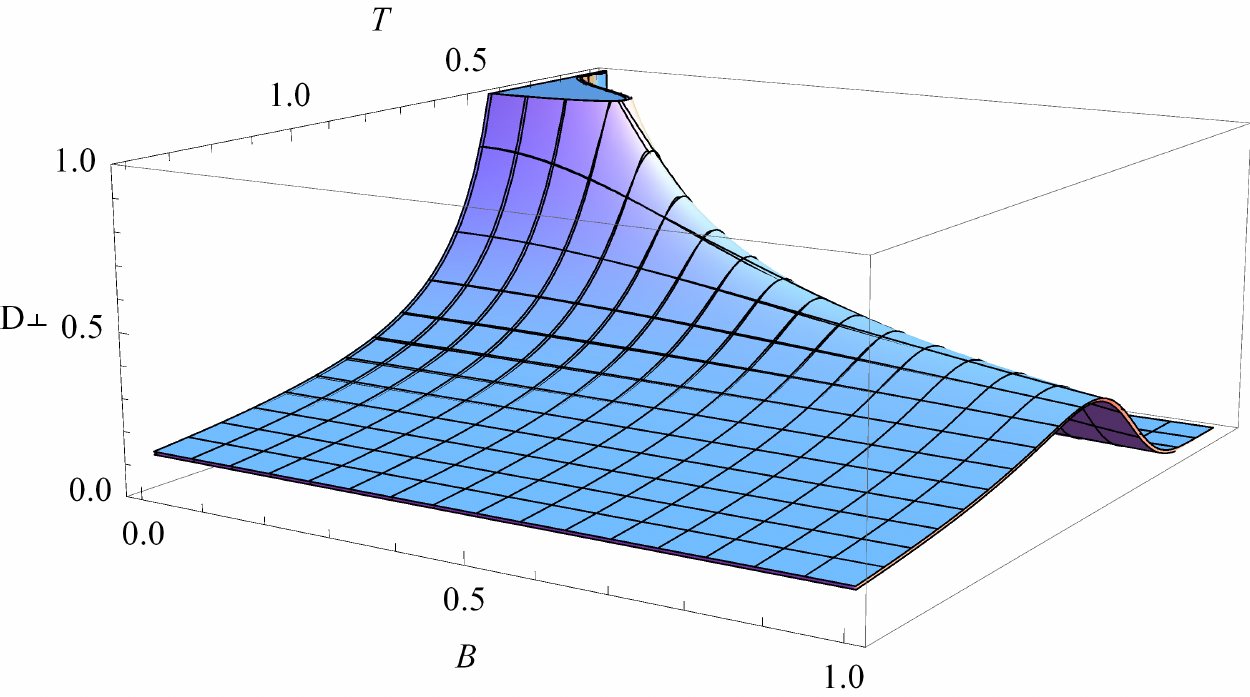}
\caption{Transverse quark diffusion constant as a function of both $B$ and $T$ using the hanging string method for $\alpha'/L^2 = 1/4$.}
\label{p3d}
\end{figure}

For $B\neq0$, the qualitative shape of $D(T,B)$ agrees with the previous spectral function approach, and is shown in Figure~\ref{p3d}.

As a conclusion on this comparison, when $B\neq 0$, the results of the spectral approach and the hanging string method are in qualitative agreement, while for $B=0$, the results differ. Our spectral function approach makes the $B=0$ regime connect smoothly to the $B\neq0$ regime, while the hanging string method does not have this feature.

\section{Indirect (medium) effects on anisotropic diffusion in a magnetic field}\label{sect6b}
So far, we only took into account the direct effect of the magnetic field ---via its coupling to the charged heavy quark--- on the diffusion. As already mentioned in the introduction, the magnetic field will also affect the medium the heavy quark thermalizes in. It would hence be instructive if we could include such indirect effect as well in our methodology. Therefore, we will stretch the original philosophy of the original soft wall model to the magnetic case: we introduce by hand a dilaton on top of the magnetic AdS$_5$ metric. For our current purposes, we are only interested in the black hole case. That metric, at leading order in $B$, was explicitly constructed by D'Hoker and Kraus in \cite{D'Hoker:2009mm,D'Hoker:2009ix} and reads
\begin{equation}
\label{bhmetric}
ds_{\text{bh}}^2=\frac{L}{z^2}\left(-f(z)dt^2+q(z)dx_3^2+h(z)\left(dx_1^2+dx_2^2\right)+\frac{dz^2}{f(z)}\right)+\mathcal{O}(B^4),
\end{equation}
with the form factors given by
\begin{eqnarray}
f(z)&=&1-\frac{z^4}{z_h^4}+\frac{2}{3}\frac{B^2z^4}{L^2}\ln \left(\frac{z}{\ell_d}\right)+\mathcal{O}(B^4), \\
q(z)&=&1+\frac{8}{3}\frac{B^2}{L^2}\int_{+\infty}^{1/z}dx \frac{\ln(z_h x)}{x^3\left(x^2-\frac{1}{z_h^4x^2}\right)}+\mathcal{O}(B^4), \\
h(z)&=&1-\frac{4}{3}\frac{B^2}{L^2}\int_{+\infty}^{1/z}dx \frac{\ln(z_h x)}{x^3\left(x^2-\frac{1}{z_h^4x^2}\right)}+\mathcal{O}(B^4).
\end{eqnarray}
The function $f(z)$ contains an extra length parameter $\ell_d$ that is a priori a completely independent scale in the problem: for any choice of $\ell_d$, Einstein's equations of motion are fulfilled with a constant magnetic field up to order $B^2$. Though, as already discussed in \cite{Dudal:2015wfn}, $\ell_d$ drops out at the end as it can be reexpressed in terms of the black hole horizon location $Z_h$ (see also later). The factor of $z_h$ in $\ln(z_h x)$ is chosen such that no singularity is encountered at $z=z_h$. \\

A schematic overview of what we will do next is presented in the bottom line of Table~\ref{tabel1}. At the end, we will combine and compare the direct and indirect effects.
\begin{table}
\begin{tabular}{|c|l|l|}
  \hline
   &  \textbf{Spectral function (second quantization)} & \textbf{Hanging string (first quantization)}\\
   \hline
  \textbf{Direct} & AdS + DBI soft wall & Lorentz force on boundary + soft wall \\\hline
  \textbf{Indirect} & D'Hoker-Kraus metric + soft wall & D'Hoker-Kraus metric + soft wall \\
  \hline
\end{tabular}
\caption{We distinguish the direct (Lorentz force) and indirect (thermal medium) influence of the magnetic field. Moreover, a second quantized spectral approach or a first quantized suspended string approach is possible, leading to 4 different calculations.}
\label{tabel1}
\end{table}

\subsection{Spectral method in D'Hoker-Kraus metric}

The temperature is related to the horizon location $Z_H$ and the magnetic field as \cite{Dudal:2015wfn}
\begin{equation}
T = \frac{1}{4\pi}\left|\frac{4}{Z_H}-\frac{2}{3}\frac{B^2}{L^2}Z_H^3\right|,
\end{equation}
which can be inverted as a power series in $B$:
\begin{equation}
Z_H = \frac{1}{\pi T} - \frac{1}{6\pi^5T^5}\frac{B^2}{L^2} + \mathcal{O}(B^4).
\end{equation}
We note that the physical magnetic field $q\mathcal{B}$ is given by $\frac{q\mathcal{B}}{1.6} = \frac{B}{L}$, as discussed in \cite{Dudal:2015wfn} by imposing a physical QCD normalization for the boundary magnetic field. The computations of $\chi$ and $D$ are the same as in Sections \ref{sect4} and \ref{sect5}, but taking the above metric \eqref{bhmetric} instead. This leads to the heavy quark susceptibility:
\begin{equation}
\chi = \frac{N_c}{6\pi^2}\frac{1}{\int_{0}^{Z_H}dz z e^{cz^2}} = \frac{N_c}{6\pi^2} \frac{2c}{e^{cZ_H^2}-1}.
\end{equation}
The validity of the perturbation series in $B$ is clearly visible here, as $\chi$ becomes divergent when $\mathcal{B} = 1.6 \sqrt{6}\pi^2 T^2$, requiring $B$ to be (much) smaller than this. The correctly normalized diffusion coefficients can then be obtained as
\begin{align}
\chi D_{\parallel} &= \frac{N_c}{6\pi}\frac{e^{-cZ_H^2}}{Z_H \pi}\frac{1}{q(Z_H)}, \\
\chi D_{\perp} &= \frac{N_c}{6\pi}\frac{e^{-cZ_H^2}}{Z_H \pi}\frac{1}{h(Z_H)},
\end{align}
with the metric warp functions expanded in $B$ as:
\begin{align}
q(Z_H) &= 1 + \frac{8}{3}\frac{B^2}{L^2}\int_{+\infty}^{\pi T}dx\frac{\ln(\frac{x}{\pi T})}{x^3(x^2-\frac{\pi^4T^4}{x^2})} + \mathcal{O}(B^4) = 1 - \frac{8}{3}\frac{B^2}{L^2}\frac{\pi^2}{96(\pi T)^4} + \mathcal{O}(B^4), \\
h(Z_H) &= 1 - \frac{4}{3}\frac{B^2}{L^2}\int_{+\infty}^{\pi T}dx\frac{\ln(\frac{x}{\pi T})}{x^3(x^2-\frac{\pi^4T^4}{x^2})} + \mathcal{O}(B^4) = 1 + \frac{4}{3}\frac{B^2}{L^2}\frac{\pi^2}{96(\pi T)^4} + \mathcal{O}(B^4). 
\end{align}
This already leads to
\begin{equation}\label{res}
\frac{D_{\parallel}}{D_{\perp}} = 1 + \frac{1}{24 \pi^2 T^4} \frac{B^2}{L^2} = 1 + \frac{1}{24 \pi^2 1.6^2} \frac{(q\mathcal{B})^2}{T^4},
\end{equation}
very analogous to the direct effect \eqref{f2}. This ratio is independent of $c$, so it equally holds in $\mathcal{N}=4$ SYM, but with a different link between $B$ and $q\mathcal{B}$ as given in \cite{D'Hoker:2009ix}. Note that the signs in the functions $q$ and $h$ are crucial to obtain this behavior. The result \eqref{res} is fully compatible with the static quark limit of the generic analysis presented in \cite{Giataganas:2013hwa}. \\

We have consistently dropped throughout a 3-determinant $\sqrt{G}\sim\mathcal{O}(B^4)$ containing only the metric in the $1,2,3$ directions, after stripping the $L^2/z^2$ overall prefactors that is. Eventually, one finds
\begin{align}
\label{indiDpa}
D_{\parallel} = \frac{1-e^{-cZ_H^2}}{2 c Z_H}\frac{1}{q(Z_H)},
\end{align}
\begin{align}
\label{indiDpe}
D_{\perp} = \frac{1-e^{-cZ_H^2}}{2 c Z_H}\frac{1}{h(Z_H)}.
\end{align}
This is formally the product of the earlier $B=0$ result (\ref{anab0}), multiplied by the (inverse) warp factor $q(Z_H)$ or $h(Z_H)$. This can be understood by rescaling the transverse coordinates on the horizon as e.g. $x^i_{new} = \sqrt{q_i(Z_H)}x^i_{old}$ for which $x^i_{new}$ is of the standard form and whose $D=D_{new}$ was hence determined earlier. Then noting that $D_i \sim \left\langle x^ix^i\right\rangle$, one obtains $D_{old} = \frac{D_{new}}{q_i(Z_H)}$ indeed. \\
At lowest order in the $B$-expansion, including the DBI soft wall \eqref{DBI} in the D'Hoker-Kraus metric merely amounts to summing both effects (as the DBI reduces to Maxwell at lowest order, the EOMs are still satisfied to the relevant order in $B$); alternatively one multiplies the above diffusion coefficients by the square root factor we considered earlier.

\subsection{Hanging string method in D'Hoker-Kraus metric}
The near-horizon expansion starts with
\begin{equation}
\phi \sim \left(1-\frac{z}{Z_h}\right)^{\pm \frac{i \omega}{\left|f'(Z_h)\right|}}
\end{equation}
and the low-frequency expansion of the field contains
\begin{equation}
\phi = B_1 + B_2 \int_{0}^{z}du \frac{u^2e^{cu^2}}{L^2q(u)f(u)}
\end{equation}
for the parallel case, and a similar expression for the perpendicular case with $q(u)$ replaced with $h(u)$.

The remainder of the derivation is completely analogous with the salient point being that the important contribution comes from differentiating this expression thrice (the $\sim \rho^3$ contribution). In the end this again leads to the coefficient $B=1 + \mathcal{O}(\nu)$. \\
The near-horizon integral over the KG inner product does contain a factor proportional to $1/f'(Z_H)$, but
\begin{equation}
f'(Z_H) = - \frac{4}{Z_H} + \frac{2}{3}\frac{\mathcal{B}^2 }{1.6^2}Z_H^3 + \mathcal{O}(\mathcal{B}^4) = - \frac{4}{Z_H} + \mathcal{O}(\mathcal{B}^4)
\end{equation}
and it is hence the same with and without $B$-field, up to the relevant order. \\
\begin{equation}\label{KG2}
(f,g) = - \frac{i}{2\pi\alpha'}\int dz \sqrt{-g}e^{-\Phi}\left|g^{00}\right|\frac{L^2}{z^2}q^I\left(f \partial_0 g^{*} - \partial_0 f g^{*}\right).
\end{equation}
Finally, we obtain:
\begin{equation}
\label{DpaHK}
D_\parallel = D_{c=0}e^{cZ_H^2} = \frac{2\alpha' Z_H}{ L^2}e^{cZ_H^2}\frac{1}{q(Z_H)},
\end{equation}
\begin{equation}
\label{DpeHK}
D_\perp = D_{c=0}e^{cZ_H^2} = \frac{2\alpha' Z_H}{L^2}e^{cZ_H^2}\frac{1}{h(Z_H)},
\end{equation}
which hence feels the magnetic field due to the shift in the horizon location. \\
Since $Z_H$ shrinks with increasing $B$, transverse diffusion decreases with increasing magnetic field. Longitudinal diffusion has two competing effects, as before. \\
These equations differ from the spectral method \eqref{indiDpa} and \eqref{indiDpe} in general, but they approach one another when $T$ becomes larger than the scale set by $c$, provided $\alpha'/L^2=1/4$. If $c=0$, then both formulas manifestly agree everywhere, provided $\alpha'/L^2=1/4$. \\
This expression is not useable to analyze what happens as $T\to 0$ when including the magnetic field, as the perturbative expansion breaks down beforehand.

\subsection{$\mathcal{N}=4$ SYM and the hanging string method}
Setting $c=0$, one finds the SYM model again. As in this case, the D'Hoker-Kraus metric is an exact dual one, it is worthwhile to compare the direct with indirect influence of the magnetic field in this case, and test the argument of \cite{Fukushima:2015wck} that the indirect diffusion should be more important. \\
The direct diffusion coefficients are given by Brownian diffusion as in \eqref{check}:
\begin{equation}
D_\perp = \frac{2\pi \alpha'L^2T^3 }{4\alpha'^2B^2 + \pi^2 L^4 T^4} = \frac{2 \alpha' }{\pi L^2 T} - \frac{\alpha'^3}{L^6}\frac{8}{\pi^3 T^5}B^2 + \mathcal{O}(B^4), \quad D_\parallel = 0,
\end{equation}
whereas the indirect effect yields (setting $c=0$ in \eqref{DpaHK}, \eqref{DpeHK}):
\begin{equation}D_\perp = \frac{2\alpha'}{\pi L^2 T} - \frac{1}{36\pi^3T^5}\frac{\alpha'}{L^2}\frac{(q\mathcal{B})^2}{1.6^2} + \mathcal{O}(\mathcal{B}^4), \quad
D_\parallel = \frac{2\alpha'}{\pi L^2 T} + \frac{1}{18\pi^3T^5}\frac{\alpha'}{L^2}\frac{(q\mathcal{B})^2}{1.6^2} + \mathcal{O}(\mathcal{B}^4).
\end{equation}
As the direct effect contains additional (inverse) factors of the 't Hooft coupling $\frac{1}{\sqrt{\lambda}} = \frac{\alpha'}{L^2}$ which in $\mathcal{N}=4$ SYM is assumed small, the indirect effect is indeed parametrically much larger here, by choice of strongly coupled regime where holography is trustworthy. \\

We will refrain from doing such an explicit comparison in the dual model(s) of QCD, as no immediate dictionary entry exists in this case, clouding the exact equivalence between the spectral and stringy approach to diffusion. The ratio $\frac{\alpha'}{L^2}$ has to be determined case by case, as we did here. Although this might appear unsatisfactory from the quantitative viewpoint (which it is), it is at least reassuring that both independent studies of the diffusion coefficients do match qualitatively.


\section{Discussion}\label{sect7}
In this paper, we continued the research started in \cite{Dudal:2014jfa} about how charmonium will react to the introduction of a strong magnetic field. We utilized a relatively simple holographic QCD model, viz.~the soft wall model, suitably adapted to take into account the coupling of the magnetic field to the constituent charm quarks. We analyzed the charm number susceptibility and (position) diffusion constants, which naturally split into a transverse and longitudinal component relative to the applied magnetic field. Our findings support a more distinct diffusion along the magnetic field direction compared to the perpendicular ones and hence strongly coupled heavy quarks in strong magnetic fields behave qualitatively the same as in the weakly coupled description: their diffusion perpendicular to the magnetic field is obstructed, and the particles hence spiral around the magnetic field lines, just like classical particles are expected to do.
From the susceptibility, we were also able to derive a new estimate for the deconfinement transition temperature, obtaining further evidence for the inverse magnetic catalysis phenomenon. Comparisons with free-field (thanks to asymptotic freedom kicking in) results were also made.

We first used the real time AdS/CFT prescription to obtain the aforementioned transport coefficients, in a second part of the paper we corroborated these from an independent methodology, based on a modeling of a heavy quark via a hanging string in the soft wall bulk. The agreement was qualitative and we did not refrain from discussing the differences between the two approaches.

The two methods we presented, the spectral function approach vs.~the hanging string method, were dealt with using a separate string length $\alpha'/L^2$. To our understanding, there is no a priori proof why both methods should give the same diffusion constants. In practice, this can only be firmly tested for models with an exactly known string dual.  Quite surprisingly, even for $\mathcal{N}=4$ SYM, there is a striking mismatch, namely one finds
\begin{equation}\label{N4}
  D_{spectral}^{\mathcal{N}=4} = \frac{1}{2\pi T}\,,\qquad D_{string}^{\mathcal{N}=4} = \frac{2}{\pi \sqrt{\lambda}T},
\end{equation}
see for example \cite[(4.2)]{Kovtun:2003wp} and \cite[(3.12)]{deBoer:2008gu}. Both results disagree unless, again, $\alpha'/L^2=1/4$. We do not have much insights to offer as to why there is an in principle very large discrepancy between the two results \eqref{N4} given the presence of the assumedly small 't Hooft coupling $\lambda$. Although one can argue that the stringy computation is only valid for extremely heavy quarks, making it non-relativistic \cite{deBoer:2008gu}, the spectral approach is not suffering from this drawback. Also, to our understanding, the stringy computation a priori assumes a Langevin description, since the diffusion constant is read off from the mean displacement squared, \eqref{displ}.
In the spectral approach, one does not have to rely on Langevin assumptions and the as such obtained transport coefficients can in principle be used to test the validity of a Langevin description for a strongly interacting quark-gluon plasma. Let us remind here that the Kubo (spectral) formalism is also what is used at the level of lattice QCD simulations to extract transport coefficients.

For the current QCD wall analysis, we must be pragmatic and hence the matching procedure from where we can start the analysis for non-zero $B$.   This naturally leads to a relatively small $\alpha'$ value. As a pro, this is compatible with general intuition from the gauge-gravity correspondence where $\frac{\alpha'}{L^2}=\frac{1}{\sqrt{\lambda}}=\frac{1}{\sqrt{g^2N}}$ is supposedly small, this to suppress stringy corrections to the geometry.  As a con, it is known that considerably larger values for $\alpha'$ are sometimes required to get reasonable values for the Polyakov (or Wilson) loop when matching these to their (lattice) QCD counterparts, see e.g.~\cite{Andreev:2006ct,Andreev:2006nw,Noronha:2009ud,Dudal:2017max}. The other standard AdS/CFT parameter constraint, $N\gg 1$ to suppress the quantum corrections, is also ``stretched'' when considering QCD with $N=3$.

We emphasize again that our model is definitely not perfect, but it does seem to provide qualitative guidelines that more sophisticated models could analyze more deeply. In that regard, the model should be thought of as a natural extension of the original soft wall model to include magnetic fields within the same phenomenological setting.
Combining the results here with those in our previous work, we believe we have demonstrated that the results obtained with this model are indeed in plausible agreement with what one would expect from QCD. Several alleys for improvement are open. We already mentioned in the introduction the (much harder to handle) influence the magnetic field can have on the medium itself and thence the indirect influence on heavy quark motion in that medium. In the current holographic model paper, we remedied that for by including magnetic field effects in the dual metric by following the magnetized soft wall setup of \cite{Dudal:2015wfn}. An improvement would consist out of constructing a self-consistent wall model with magnetic field included. The latter would probably also allow a more profound comparison between the spectral and hanging string method if the confinement behavior is properly present in the metric without having to jeopardize the underlying gravitational equations of motion. Such set-up might also be helpful to get a better handle on how to treat/fix the string length (or $\alpha'$) in relation to QCD parameters. Another point where things can be improved is in the description of the holographic charmonium properties, next to the mass also the (thermal) width etc, when compared to QCD, as discussed in \cite{Grigoryan:2010pj,Braga:2018zlu}.\footnote{In \cite{Braga:2018zlu}, an improved phenomenological dilaton ansatz was imposed on top of the D'Hoker--Kraus metric, while maintaining the Maxwell--Einstein--Hilbert action. This corresponds to a study of the indirect medium effects, as the direct coupling of the magnetic field to the charged constituents is ignored.} Work along these lines is in progress and will be reported about in the future.

\section*{Acknowledgments}
We thank D.~Giataganas for interesting comments, next to D.~R.~Granado for discussions when this project was initiated. T.~G.~Mertens gratefully acknowledges financial support from the Research Foundation-Flanders (FWO Vlaanderen) next to the Fulbright program and a Fellowship of the Belgian American Educational Foundation during the early stages of this research.

\appendix
\section{Diffusion in an anisotropic medium}
\label{formulas}
We provide here some details about the diffusion coefficients in anisotropic media. \\
Firstly, to lowest order in density fluctuations, diffusion is governed by Fick's second law,
\begin{equation}
\partial_t n = \nabla \left(\cdot \mathfrak{D} \cdot \nabla n\right),
\end{equation}
where $\mathfrak{D}$ is a $3\times 3$ rank 2 tensor: the diffusion matrix. \\

By definition, the retarded Green function of the conserved currents is given by
\begin{equation}
G^{R}_{\mu\nu}(\omega, \mathbf{k}) := -i \int dt \int dV e^{i\mathbf{k}\cdot \mathbf{x}}e^{-i\omega t}\theta(t)\left\langle \left[J_{\mu}(t,\mathbf{x}),J_{\nu}(0,\mathbf{0})\right]\right\rangle_{\beta}.
\end{equation}
Combining linear response theory with the above anisotropic diffusion equation, one obtains analogously as in the isotropic case \cite{Laine:2016hma, pasztor} the following link between the retarded Green function and the diffusion matrix:
\begin{equation}
G^R_{00} (\omega, \mathbf{k}) = -\chi \left( \mathbf{k}\cdot \mathfrak{D} \cdot \mathbf{k}\right)\frac{\left( \mathbf{k}\cdot \mathfrak{D} \cdot \mathbf{k}\right) + i\omega}{\omega^2 + \left( \mathbf{k}\cdot \mathfrak{D} \cdot \mathbf{k}\right)^2}.
\end{equation}
Taking the real part of this equation, one immediately finds agreement with expression (\ref{qnsus}).
Current conservation $\partial_\mu J^\mu = 0$ of the currents allows this expression to be rewritten as
\begin{equation}
\label{ani}
\frac{k^l G^R_{lm} k^m}{\omega^2} = -\chi \left( \mathbf{k}\cdot \mathfrak{D} \cdot \mathbf{k}\right)\frac{\left( \mathbf{k}\cdot \mathfrak{D} \cdot \mathbf{k}\right) + i\omega}{\omega^2 + \left( \mathbf{k}\cdot \mathfrak{D} \cdot \mathbf{k}\right)^2}.
\end{equation}
As this expression must hold for any sufficiently small $\mathbf{k}$, one immediately reads off that
\begin{equation}\label{A5}
\mathfrak{D}_{ij} = -\lim_{\omega \to 0} \frac{1}{\chi} \frac{\Im G^R_{ij}(\omega, \mathbf{0})}{\omega}.
\end{equation}
In our specific case, the diffusion matrix is anisotropic but diagonal due to a lack of off-diagonal coupling terms in the Lagrangian, and we denote $D_i := \mathfrak{D}_{ii}$.
Note finally that as long as $\omega \neq 0$, the point $\mathbf{k}=0$ is a regular point of the function $G^R_{ij}(\omega, \mathbf{k})$, as the above equality (\ref{ani}) demonstrates, implying that no funny business happens as one changes the ``direction of approach'' in taking the $\mathbf{k} \to \mathbf{0}$ limit. Of course, this \emph{does} happen when swapping the $\omega \to 0$ limit with the $\mathbf{k}\to \mathbf{0}$ limit.

\end{document}